\documentclass[fleqn,usenatbib]{mnras}

\usepackage[T1]{fontenc}
\usepackage{ae,aecompl}
\usepackage{algorithm}
\usepackage{algorithmic}

\DeclareRobustCommand{\VAN}[3]{#2}
\let\VANthebibliography\thebibliography
\def\thebibliography{\DeclareRobustCommand{\VAN}[3]{##3}\VANthebibliography}

\usepackage{graphicx}	% Including figure files
\usepackage{amsmath}	% Advanced maths commands
\usepackage{natbib}
\usepackage{bm}         % bold face math symbols
\usepackage{xspace}
\usepackage{xcolor}
\usepackage{verbatim}

\usepackage{mwe}
\usepackage{subcaption}
\usepackage{newtxtext,newtxmath}

\newcommand{\synz}{\textsc{syth-z}\xspace}
\newcommand{\umachine}{\textsc{universemachine}\xspace}
\newcommand{\fsps}{\textsc{fsps}\xspace}
\newcommand{\illustris}{IllustrisTNG\xspace}

\newcommand{\Mpeak}{\ensuremath{M_\mathrm{peak}}\xspace}
\newcommand{\mahslope}{\ensuremath{\beta_\mathrm{MAH}}\xspace}

\title[Synthetic spectra for redshift estimation: {\rm{SYTH-Z}}]{Machine learning synthetic spectra for probabilistic redshift estimation: {\rm{SYTH-Z}} }

\author[Ramachandra et al.]{Nesar Ramachandra $^{1, 2}$\thanks{E-mail:nramachandra@anl.gov},
Jon\'as Chaves-Montero$^{2,3}$,
Alex Alarcon$^{2}$,
Arindam Fadikar$^{4}$, 
Salman Habib$^{1, 2}$,
\newauthor
Katrin Heitmann$^{2}$
\\$^{1}$ Computational Science Division, Argonne National Laboratory, 9700 South Cass Avenue, Lemont, IL 60439, USA.
\\$^{2}$ High Energy Physics Division, Argonne National Laboratory, 9700 South Cass Avenue, Lemont, IL 60439, USA.
\\$^{3}$ Donostia International Physics Center, Paseo Manuel de Lardizabal 4, 20018 Donostia-San Sebastian, Spain.
\\$^{4}$ Mathematics and Computer Science Division, Argonne National Laboratory, 9700 South Cass Avenue, Lemont, IL 60439, USA.
}

\date{Accepted XXX. Received YYY; in original form ZZZ}

\pubyear{2021}

\begin{document}

% \tableofcontents

\label{firstpage}
\pagerange{\pageref{firstpage}--\pageref{lastpage}}
\maketitle

% Abstract of the paper
\begin{abstract}
Photometric redshift estimation algorithms are often based on representative data from observational campaigns. Data-driven methods of this type are subject to a number of potential deficiencies, such as sample bias and incompleteness. Motivated by these considerations, we propose using physically motivated synthetic spectral energy distributions in redshift estimation. In addition, the synthetic data would have to span a domain in colour-redshift space concordant with that of the targeted observational surveys. With a matched distribution and realistically modelled synthetic data in hand, a suitable regression algorithm can be appropriately trained; we use a mixture density network for this purpose. We also perform a zero-point re-calibration to reduce the systematic differences between noise-free synthetic data and the (unavoidably) noisy observational data sets. This new redshift estimation framework, \synz, demonstrates superior accuracy over a wide range of redshifts compared to baseline models trained on observational data alone. Approaches using realistic synthetic data sets can therefore greatly mitigate the reliance on expensive spectroscopic follow-up for the next generation of photometric surveys.
\end{abstract}


%%%%%%%%%%%%%%%%%%%%%%%%%%%%%%%%%%%%%%%%%%%%%%%%%%

%%%%%%%%%%%%%%%%% BODY OF PAPER %%%%%%%%%%%%%%%%%%

\section{Introduction}
\label{sec:intro}
 
Ongoing and near-future photometric surveys such as the Dark Energy Survey \citep[DES,][]{Abbott:05}, the Kilo-Degree Survey \citep[KiDS,][]{de_Jong:13}, the Hyper Suprime-Cam Survey \citep[HSC,][]{Aihara:2018a,Aihara:2018b}, Euclid \citep{Laureijs:11}, the Vera C. Rubin Observatory's Legacy Survey of Space and Time~\citep[LSST,][]{Abell:09}, and the Roman Space Telescope~\citep{Green:12} encounter several hurdles that were not faced by previous, more shallow, surveys such as the Sloan Digital Sky Survey \citep[SDSS,][]{York:00}. As galaxy catalogues continue to be substantially larger in size and fainter in magnitude, spectroscopic follow-ups to obtain confirmation of galactic properties become prohibitively expensive \citep{Newman:2015, Stanford2021}. This necessitates alternatives to spectroscopic follow-ups via synthetic modeling, using a combination of stellar population synthesis (SPS) models and star formation histories (SFHs) from parametric models \citep[e.g.,][]{alsing2020_SPECULATOREmulatingstellar}, semi-analytic models \citep[e.g.,][]{pacifici2012_RelativeMeritsdifferent}, or hybrid approaches combining empirical and semi-analytic models \citep[e.g.,][]{dc2}. 
 
Photometric redshift estimation of individual galaxies typically relies upon data either in the form of Spectral Energy Distributions (SED) templates for matching \citep{Fernandezsoto:99, Benitez:00}, or as training sets for machine learning algorithms \citep{Firth:03}. The supervised learning algorithms are designed to find a mapping between sets of dependent variables based on a training set with input-target examples. In an ideal case, these algorithms can predict outputs (like galaxy redshifts) for combinations of input variables (like galaxy colours) never previously encountered, therefore generalizing training data to unseen situations. However, in a more realistic situation, supervised learning algorithms inherit biases presented in the training data. Therefore, to prevent these biases, it is mandatory to optimise supervised learning models using training sets encompassing all cases expected to be found in the envisaged application scenario. 

The majority of machine learning (ML) algorithms designed to infer photometric redshifts use observational photometric data for training \citep{Carrasco2013, Carrasco2014, Graff2014, Almosallam2016a, Almosallam2016b, Cavuoti2016, Sadeh2016, Graham2018, fadikar2021scalable}. This approach is not ideal, however, because spectroscopic surveys can only estimate redshifts at the bright end of sources detected by photometric surveys, given their significantly slower observing speeds. Furthermore, aiming to reduce the level of star-galaxy misclassification, most spectroscopic surveys incorporate strict selection criteria on colour space that result in not being able to observe some galaxy types. Taken together, these issues may lead to unrealistic predictions for galaxy types, especially faint galaxies with uncommon properties. One approach to overcome biases originating from incomplete observations is to employ stellar population synthesis models to generate more complete training data sets. 

The viability of stellar population synthesis modelling in photometric redshift estimation is currently explored in template fitting codes. For instance, \citet{hernan-caballero2021_MiniJPASSurveyPhotometric} use the stellar population synthesis code \textsc{cigale} \citep{Boquien2019} with delayed-exponential star formation histories to create templates, and then introduce these in a customized version of the template-fitting code \textsc{lephare} \citep{Arnouts:99} to compute photometric redshifts for galaxies of the miniJPAS survey \citep{bonoli2021_MiniJPASSurveypreview}, which observed $\simeq1\deg^2$ of the northern sky using the 56 narrow-band filters of the J-PAS survey \citep{benitez2014_JPASJavalambrePhysicsAccelerated}. A similar approach is also utilized in the COSMOS-2020 \citep{Weaver2021} catalogue or the PAUS survey \citep{PAUSCOSMOSphotoz}, where log-normal star formation histories are used to create 17 generated synthetic SEDs from Flexible Stellar Population Synthesis (\fsps, \citealt{Conroy2009, conroy_gunn_2020}). These are also utilized in the template fitting code \textsc{eazy} \citep{Brammer2008} to estimate the redshift. Both these approaches, while demonstrating the usefulness of synthetic SEDs, also highlight the need for more realistic galaxy formation models in the future surveys of LSST, Euclid and SPHEREx \citep{spherex}. Furthermore, expanding the template set to include more realistic and diverse galaxy formation physics often leads to an unaffordable increase in compute time for current and future surveys.
%These approaches are also affected by photometric errors, especially for fainter magnitudes. 

Another issue with photometric redshift estimation is that the mapping is not an injective function, due to the relatively sparse input dimension, resulting in degeneracies. Hence a simple point-estimate for a galaxy's ``photo-z'' is less desirable than a probability distribution function (PDF) that is conditioned on the observed photometric colours, i.e, $p(z|c)$~\citep{mandelbaum:2008, Almosallam2016b, Benitez2000, Izbicki:17b, Carrasco_Kind:13}, and possibly other properties, such as morphology and galaxy clustering. However, a full Bayesian posterior estimation of $p(z|c)$ may not be practically feasible with traditional statistical approaches. This is primarily due to nontrivial issues with scaling the inference problem to billions of galaxies. Despite having a robust forward model and massive parallelization abilities, obtaining full redshift posteriors for individual galaxies is still prohibitively expensive. 

Artificial neural networks could be used as inference machines as alternatives to traditional posterior estimation techniques, by sampling over the model parameters of network weights and biases \citep{neal2012bayesian}. However, the numerical expense of sampling over a large number of model parameters can be numerically expensive, as is the case with deep neural networks. Consequently, neural networks employing Bayesian inference techniques for robust uncertainty estimates \citep{mackay1995, gal2016uncertainty, 2020arXiv200202405W} are an active field of research. 

It should also be noted that for photometric redshift estimation and downstream analyses, a simpler conditional density estimator that quantifies the uncertainty in the predictions with respect to the data-prior suffices for most purposes. Approximating the likelihoods using parameterized models could further reduce the computational expense. This could provide the much-needed scalability to billions of galaxies, whilst providing robust uncertainty estimations for redshift predictions. Hence we adopt this approach in our photometric redshift estimation by approximating the conditional distribution $p(z|c)$ using a mixture of Gaussian distributions. Our redshift estimation algorithm \synz is constructed with both the above considerations: the synthesis of physically modelled mock photometric data, and adaptation of synthetic data in a probabilistic framework that is scalable to a large number of galaxies. In this paper, we show that with robust physical modelling and meticulous sampling, we may achieve high levels of accuracy in photo-z estimation over a baseline of observation-based training. 

The methodology described above is a {\it hybrid} approach that bridges aspects of both ML and template methods of photometric redshift estimation. The ML interpretation is straightforward -- weights of a neural network are optimized to learn the mapping between broad-band photometry and redshifts. From a template-fitting methods perspective, \synz is built upon complex empirical star formation histories (SFHs) and SEDs, and these are approximated using an interpolation technique. This novel combination makes \synz a powerful framework for predicting galactic redshifts. 

The paper is organized as follows. The process of creating the synthetic galaxy colours is introduced in Section~\ref{sec:data_syn}. Details of the data from the photometric surveys of SDSS, VIPERS and DEEP2 are described in Section~\ref{sec:data_obs}. In Section~\ref{sec:modeling} we explain the probabilistic machine learning algorithm used for computing the redshift PDF, given galaxy colours. We also introduce zero-point adjustments performed to reduce the systematic differences between synthetic and observed colours. In Section~\ref{sec:results}, we demonstrate the advantage of using synthetically-trained photometric redshift estimation against a baseline training scheme of using survey data for training. We discuss our findings and summarize the major results in Section~\ref{sec:discussion}. % and summarize our principal results in Section~\ref{sec:conclusions}.

Throughout this work, we use extinction-corrected magnitudes in the AB system and {\it Planck} 2015 cosmological parameters \citep{planck14b}: $\Omega_{\rm m}= 0.314$, $\Omega_\Lambda = 0.686$, $\Omega_{\rm b} = 0.049$, $\sigma_8 = 0.83$, $h_0 = 0.67$, and $n_{\rm s} = 0.96$.

%%%%%%%%%%%%%%%%% %%%%%%%%%%%%%%%%% %%%%%%%%%%%%%%%%% 

\section{Training and validation data}
\label{sec:data}

We begin with the curation of the training and validation data used throughout this paper. In contrast to the traditional machine learning frameworks, the testing or validation set is different from the training data. That is, the \synz training is entirely carried out with the synthetic data, and `out-of-set' data is used for benchmarking. In this section, we begin by presenting the process of modelling synthetic training data from stellar population synthesis codes. Next, we tabulate the observational galaxy samples used for validating the photo-$z$ algorithm introduced in Section \ref{sec:gmm}. 

%%%%%%%%%%%%%%%%%%%%%%%

\subsection{Synthetic spectral energy distributions}
\label{sec:data_syn}

Modelling galaxy SEDs is a long-standing problem in astronomy due to the dependence of the light of a galaxy on several properties and their correlations, such as star formation history, metallicity, dust type and abundance, and gas properties. The most common approach to predict galaxy SEDs is by using stellar population synthesis (SPS) models, which rely upon stellar evolution theory to produce precise galaxy spectral energy distributions in the ultraviolet, optical, and infrared ranges ~\citep[for a review see][]{Conroy2013}. Throughout this work, we generate galaxy colours using the SPS model~\fsps~\citep{Conroy2009, conroy_gunn_2020} together with its python interface \citep{python_fsps}, which include a plethora of parameters controlling galaxy properties. Our strategy is to produce synthetic galaxy photometry approximately spanning the same region of the colour space as galaxies from observations (see Section \ref{sec:data_obs}). 

Producing synthetic colours representative of the colours of the observed galaxies is challenging due to the large diversity of properties influencing colours. For example, \citet{Pacifici2015} show that precise modelling of star formation histories, metallicity enrichment histories, dust attenuation, and nebular emission is necessary to produce colours covering the same region of the colour space as observed galaxies. Among these properties, the star formation history presents the strongest influence on broadband colours \citep[e.g.,][]{chaves_hearin2020_sbu1}. The two main approaches for modelling SFHs are parametric functional forms and star formation rates tabulated at a set of cosmic times. Sufficiently flexible parametric models capture SFHs predicted by simulations with reasonable precision \citep{simha_etal14, diemer_etal17} and accommodate episodic bursts of star formation or other transient features. But simple models may fail to do so entirely \citep{Lower2020}. Another crucial concern of the parametric models is that they may produce SFHs that are not physically motivated. 

Motivated by the above issues, we use SFHs predicted by galaxy formation models to produce physically-motivated colours. The main drawback of this approach is that the results become model-dependent. To alleviate this problem, we draw SFHs from two models that reproduce a broad range of summary statistics of galaxy populations but follow completely different approaches to simulate galaxies. The first is the empirical model \umachine~\citep{Behroozi2019}, which simulates galaxies using a set of scaling relations between galaxy and halo properties. We generate \umachine SFHs by running the publicly available version of the code\footnote{\url{https://bitbucket.org/pbehroozi/universemachine/}} on merger trees identified in the Bolshoi-Planck simulation with Rockstar and Consistent trees~\citep{behroozi_etal13a, behroozi_etal13b, klypin_etal16, rodriguez_puebla_etal16} up to redshift $z=0$ and 1. The second is based on the cosmological hydrodynamical simulation \illustris~\citep{Marinacci2018, Naiman2018, Nelson2018b, Pillepich2018a, Springel2018}, which models the joint evolution of dark matter, gas, stars, and supermassive black holes by incorporating a comprehensive galaxy formation model with radiative gas cooling, star formation, galactic winds, and AGN feedback. In particular, we draw \illustris SFHs from the largest hydrodynamical simulation of the suite, TNG300-3~\citep{Nelson2018a}.

Due to the non-parametric nature of SFHs drawn from \umachine and \illustris, it is challenging to select a set of SFHs representative of a whole galaxy population. To facilitate this process, we label each SFH by the maximum mass ever attained by its host halo, \Mpeak, and the slope of the host halo mass accretion history, \mahslope. It is natural to use \Mpeak because galaxies hosted by more massive haloes present larger masses, reach the peak of their SFH at higher redshift, and become quenched at earlier times~\citep[e.g.,][]{chaves_hearin2020_sbu1}. We use \mahslope because this parameter captures the dependence of galaxy properties on the assembly history of its host halo; for example, at fixed halo mass, the galaxies with a steeper slope reach the peak of their SFH at higher redshift and become quenched at earlier times~\citep{MonteroDorta2021}.

We also model the impact of three other galaxy properties influencing colours: metallicity, dust attenuation, and nebular emission. We use a time-independent parameter $Z$ to model the stellar and gas metallicity, $\gamma \tau_{V}^{\rm ISM}$ and $\tau_{V}^{\rm ISM}$ to specify the dust attenuation of light coming from stars younger and older than 10 Myr~\citep{Charlot2000}, respectively, and $\log_{10} U_S$ to set the logarithm of zero-age ionization at the Str\"omgren radius. Therefore, our model presents six free parameters: $Z$ and $\log_{10} U_S$ are defined at 22 and 7 values within the ranges $Z\in[0.0002, 0.03]$ and $\log_{10} U_S\in[-4, -1]$ respectively, $\gamma$ and $\tau_{V}^{\rm ISM}$ are continuous to within the intervals $\gamma\in[1, 4]$ and $\tau_{V}^{\rm ISM}\in[0, 1.5]$, and \Mpeak and \mahslope vary to within the ranges predicted by \umachine and \illustris. We note that we also use a Chabrier initial mass function~\citep{Chabrier2003} to predict colours.

To produce galaxy colours, we begin by sampling the model parameters according to a Latin hypercube design~\citep{McKay1979} defined by the priors specified above. To reduce the impact of short-term star formation fluctuations on colours, we compute the average of the 20~SFHs with closest \Mpeak and \mahslope to the sampled values. Note that the resulting SFHs are more representative of the colours of galaxy populations~\citep{ChavesMontero2021}. Then, we use \fsps to generate colours at 50 randomly selected redshifts from $z=0.002$ to 1.25 for each combination of model parameters. In this manner, we end up producing synthetic colours for $\sim$200\,000 different galaxies. In Appendix~\ref{app:tempaltes}, we compare the spectral energy distribution of some of these galaxies and that of galaxies from observations.

%%%%%%%%%%%%%%%%% %%%%%%%%%%%%%%%%% %%%%%%%%%%%%%%%%% 

\subsection{Data from observations}
\label{sec:data_obs}

The standard approach to evaluate the quality of photometric redshifts is to use spectroscopic redshift estimates as ground truth, which is motivated by the much better precision of spectroscopic redshifts relative to photometric redshifts. Nonetheless, the precision of spectroscopic redshifts decreases significantly for low signal-to-noise sources; consequently, this technique may lead to catastrophic redshift solutions for faint sources. In this section, we describe the main properties of the galaxies with secure spectroscopic redshift estimates that are used to validate our methodology.

In order to benchmark \synz with a broadly representative sample of galaxies with spectroscopic redshifts and broadband photometry, we collate publicly available spectroscopic sources from the Sloan Digital Sky Survey~\citep[SDSS,][]{York:00, Eisenstein2011_SDSS3, Blanton2017_SDSS4}, the VIMOS Public Extragalactic Redshift Survey~\citep[VIPERS,][]{Scodeggio2018_VIPERS}, and the DEEP2 Redshift Survey~\citep[DEEP2,][]{Newman2013_DEEP2, Matthews2013_DEEP2}. We detail the properties of each of the samples below, having tabulated the primary statistics in Table~\ref{tab:obv_data}.

\begin{enumerate}

    \item {\bf SDSS: } This sample comprises the 1,911,919 galaxies from the SDSS database\footnote{SDSS data are available from: \url{http://skyserver.sdss.org/dr15/en/tools/search/sql.aspx}} presenting the spectroscopic redshifts with precision superior to $\Delta z_{\rm spec}=0.003$, extinction-corrected $i$-band magnitude brighter than 20 mag, $i$-band magnitude error smaller than 0.1 mag, and clean photometry in all bands. For these sources, we use extinction-corrected magnitudes from the public database. These galaxies present redshifts typically smaller than $z_{\rm spec}=0.6$. 
    
    \item {\bf VIPERS: } We select the 40,718 galaxies from the second public data release of the VIPERS survey\footnote{VIPERS data are available from: \url{http://vipers.inaf.it/rel-pdr2.html}} with secure redshifts. VIPERS galaxies present redshifts slightly larger than the SDSS galaxies, but typically smaller than $z_{\rm spec}=1$. 
    
    \item {\bf DEEP2: } We use the 13,163 galaxies from the fourth data release of DEEP2\footnote{DEEP2 data are available from: \url{http://deep.ps.uci.edu/DR4/home.html}} with spectroscopic redshifts. These galaxies present redshifts covering approximately the same redshift range as VIPERS galaxies.

\end{enumerate}

\begin{table}
    \centering
    \caption{Properties of the spectroscopic galaxy samples that we use for model validation. The columns i$_\mathrm{max}^{95\%}$ and $z_\mathrm{max}^{95\%}$ indicate the $i$-band magnitude and redshift of the galaxy in the 95th percentile of each of these properties.}
    \begin{tabular}{crcc}
        Survey & Number & i$_\mathrm{max}^{95\%}$ [mag] & $z_\mathrm{max}^{95\%}$ \\
        \hline
        SDSS    & 1\,911\,919 & 19.8 & 0.64\\
        VIPERS  &     40\,718 & 22.5 & 0.97\\
        % PRIMUS  &     31\,317 & 23.1 & 0.92\\
        DEEP2   &     13\,163 & 22.9 & 0.97\\
        \hline
    \end{tabular}
    \label{tab:obv_data}
\end{table}

Some galaxies from the VIPERS and DEEP2 surveys present photometric data taken by the Canada-France-Hawaii Legacy Survey~\citep[CFHTLS,][]{Gwyn2012_CFHTLS}. This survey's filter transmission curves have slightly different band-passes relative to the SDSS filters. To avoid biases arising from such differences, we transform the photometry of these galaxies from the CFHTLS into the SDSS filter system using Equations~(1-5) from~\citep{Matthews2013_DEEP2}. We also correct the magnitude of all these galaxies for Galactic extinction.

%%%%%%%%%%%%%%%%%%%%%%%
%%%%%%%%%%%%%%%%%%%%%%%

\section{Modelling and framework} \label{sec:modeling}

Unlike the majority of ML-based photo-z estimation codes, our framework is trained on noiseless, simulated data. The discrepancy in training and testing samples requires calibrations and validations beyond standard training and hold-out testing routines. In this section, we describe the major components in the \synz framework that aim to address these concerns: experimental design and synthesis of training data, construction and training of the probabilistic mapping algorithm, and survey-specific systematic offset corrections.  

\subsection{Colour distribution of observational and synthetic galaxies}

The architecture of data-driven ML methods, e.g., artificial neural networks, is usually agnostic to the underlying physics. Only the training data includes the intrinsic information about the input and target distributions, and the underlying relationships between them. Thus any incorrect modelling or biases in the training data will be learned by the network. This is particularly important when the training data is created using simulations -- the assumptions in the synthetic modelling may create intractable biases in the estimations. 

In our work, the mixture density network (MDN) learns the information about mapping between the colours and redshift entirely from the training data. Since the broadband colours for training \synz are synthetically generated via the steps prescribed in Section \ref{sec:data_syn}, the validity of synthetic data has to be meticulously checked. Primarily, the coverage of synthetic data has to be compared to that of observational samples. In the colour-redshift space, if the distribution training set does not significantly overlap with that of the observational data, the resulting extrapolation is highly prone to biases in the redshift estimations. 

\begin{figure*}
\centering
\begin{minipage}{0.95\textwidth}
    \centering
\begin{subfigure}[b]{0.99\textwidth}\centering
        \includegraphics[width=0.9\textwidth]{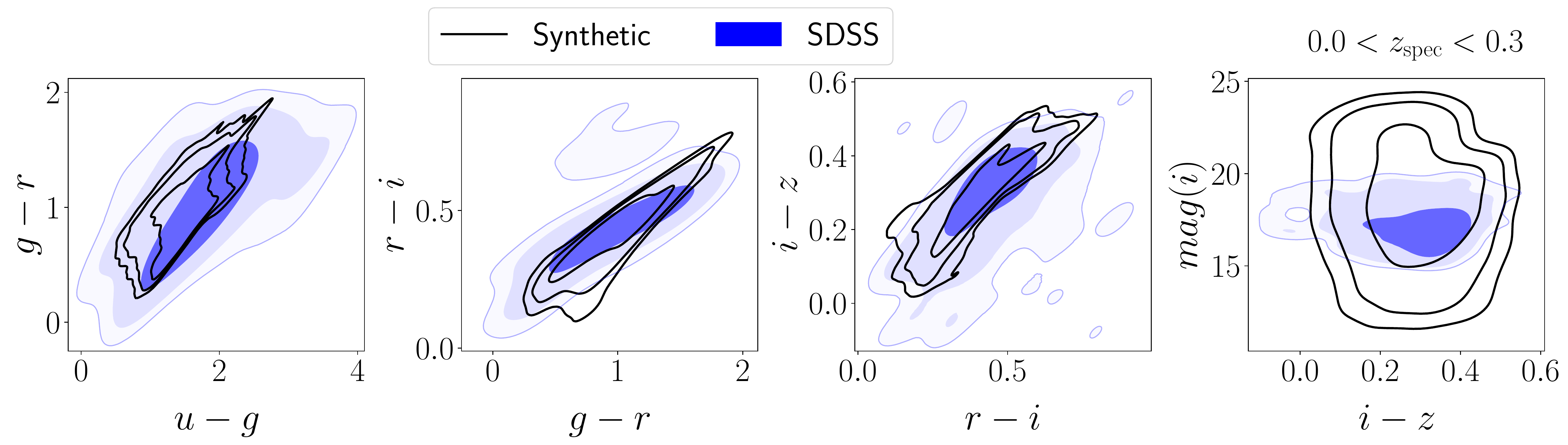}
        \caption{Comparison of the synthetic and observed (SDSS) colour distributions for galaxies in the redshift range $0.0 < z < 0.3$.}
        \label{fig:first_sub}
\end{subfigure}
    \\
\begin{subfigure}[b]{0.99\textwidth}\centering
        \includegraphics[width=0.9\textwidth]{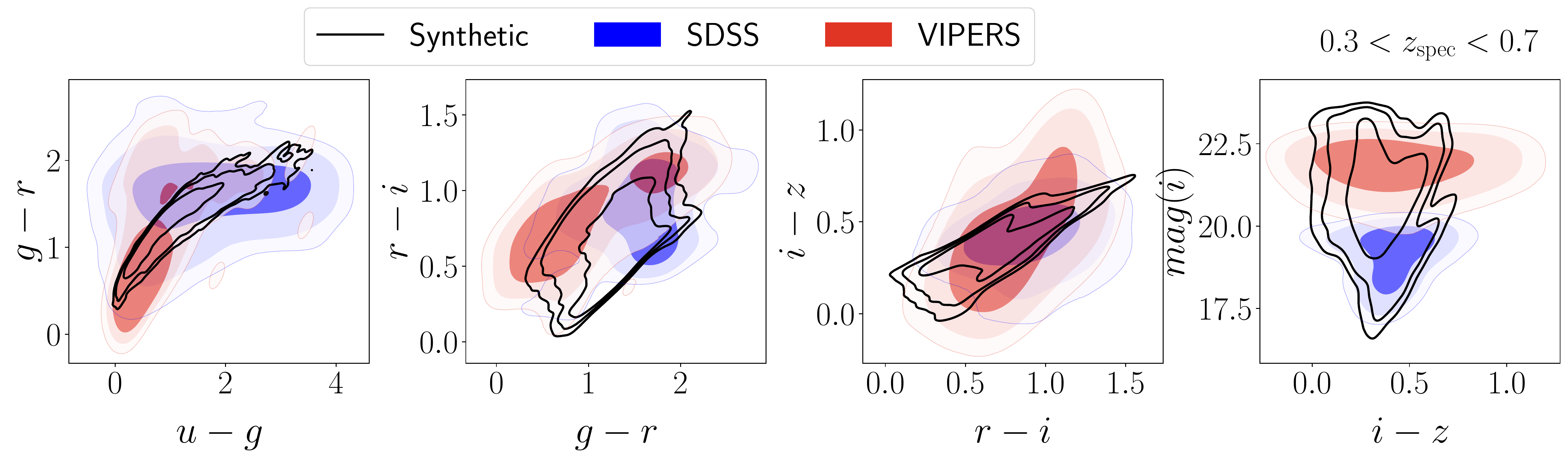}
        \caption{Comparison of the synthetic and observed (SDSS and VIPERS) colour distributions, over the redshift range $0.3 < z < 0.7$.}
        \label{fig:second_sub}
\end{subfigure}
    \\
\begin{subfigure}[b]{0.99\textwidth}\centering
        \includegraphics[width=0.9\textwidth]{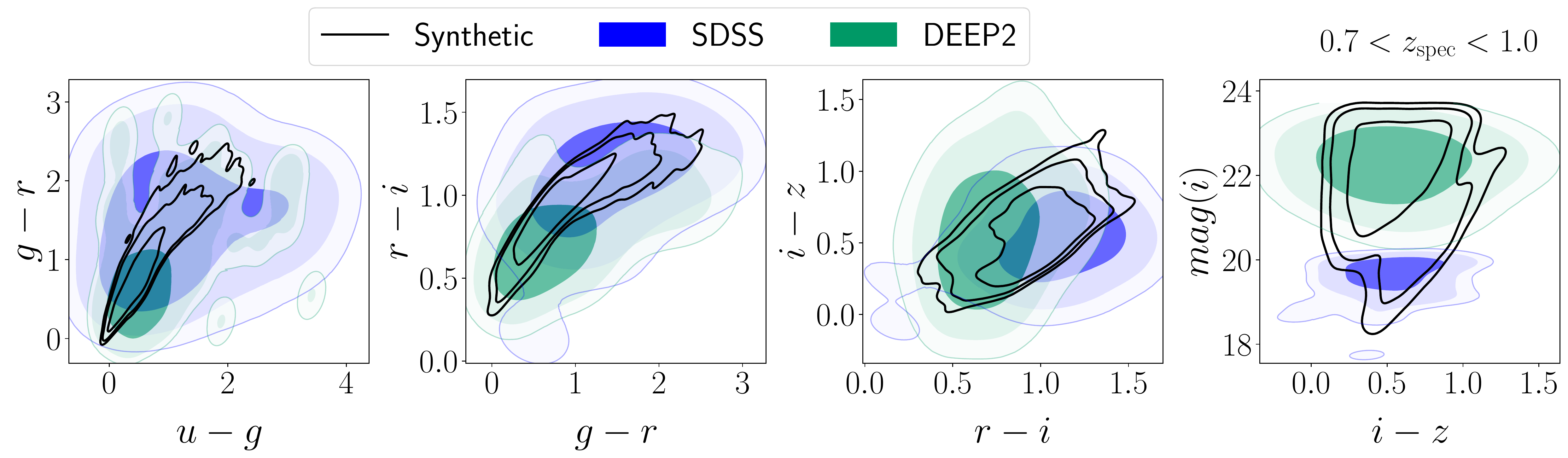}
        \caption{Colour distribution comparison following panel (b), for synthetic and observed (SDSS and DEEP2) colour distributions for galaxies in the redshift range $0.7 < z < 1.0$.}
        \label{fig:third_sub}
\end{subfigure}
    \\
\begin{subfigure}[b]{0.99\textwidth}\centering
        \includegraphics[width=0.9\textwidth]{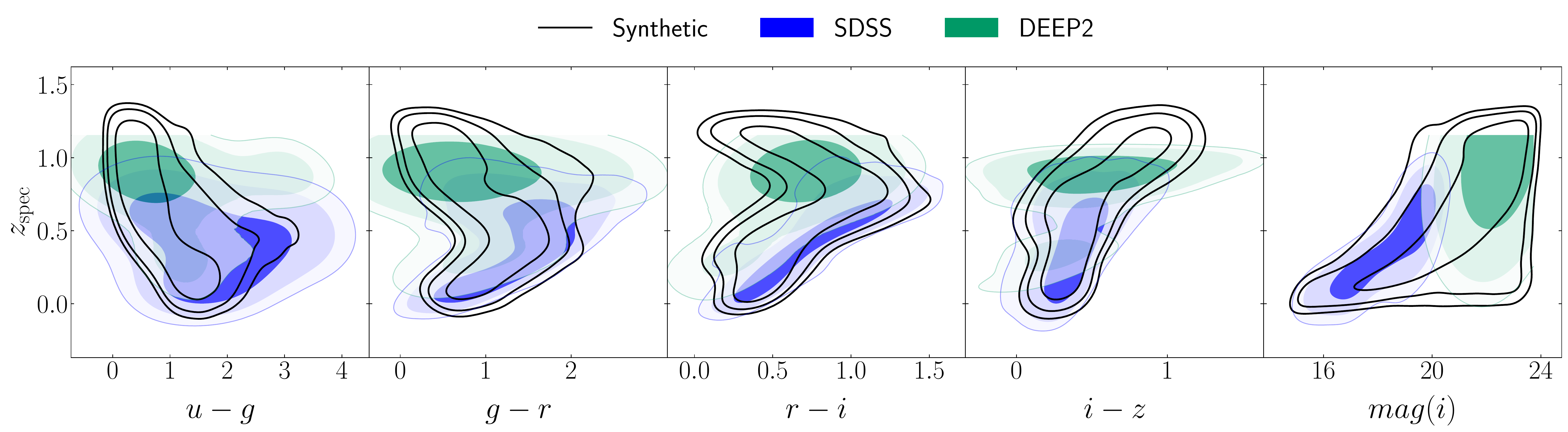}
        \caption{Distribution of galaxy colours and i-magnitude with redshift for the synthetic, SDSS and DEEP2 data sets.}
        \label{fig:fourth_sub}
\end{subfigure}

    \caption{Comparison of synthetic and observational data: Panels show different projections of the joint distributions of colours (panels (a), (b), (c)), and of the colours and i-magnitude and redshift (panel (d)).}
    \label{fig:data_colour_comparison}
\end{minipage}
\end{figure*}

In Figure~\ref{fig:data_colour_comparison}, we display the distribution of colours and spectroscopic redshifts ($u-g$, $g-r$, $r-i$, $i-z$, $i$-band, $z_{\rm spec}$) projected onto 2D slices. The first, second, and third panels show the colour distribution of observed and synthetic data at $0.0 < z_{\rm spec} < 0.3$, $0.3 < z_{\rm spec} < 0.7$, and $0.7 < z_{\rm spec} < 1.0$, respectively. We find that the distribution of SDSS colours spans the same region of the colour space as that of synthetic colours for the first redshift bin but not for the higher redshift bin of $0.7 < z_{\rm spec} < 1.0$. This is because the SDSS low-redshift sample is primarily a magnitude-limited sample, while at intermediate and high redshift the survey only observed red and bright galaxies. In fact, we can see that the distribution of SDSS colours in the second redshift bin agrees with just the distribution of colours for the reddest synthetic galaxies. 

In the third redshift bin $0.7 < z_{\rm spec} < 1.0$, we can readily see that SDSS galaxies present, on an average, redder colours than DEEP2 galaxies. The red colours of SDSS galaxies are once again explained by the flux limit selection, which at higher redshifts preferentially selects luminous red galaxies. DEEP2 selects much fainter galaxies (up to $i\sim24$), with a particular colour selection that targets galaxies with $z > 0.7$, yielding a different colour distribution with a larger population of blue galaxies. Furthermore, we can see that the distribution of galaxy colours for observed data is more extended than for the case of synthetic data. This is simply because we produce noiseless synthetic colours and observed colours suffer from significant photometric uncertainties, especially for faint galaxies. Possible systematic errors by training on such noiseless data may require corrections (as explained in Section \ref{sec:zp}) and calibrations (as explained in Section \ref{sec:margin}) when adapting to real observations.

We can also see that synthetic galaxies span the same region of the colour-redshift space as observed galaxies, which strongly suggests that our methodology produces synthetic colours representative of the observed colours of spectroscopic galaxies. This, in conjunction with physically motivated information regarding the colour-redshift correlation in the training set would enable an ML model to learn the associated mapping. Furthermore, given that observed data from different surveys do not span the same region of the colour space, an ML-model trained on one survey may not be apt for inferring redshifts from another. With a training prior different from that of testing colour-redshift space, the mapping learned could be inherently biased, and a leading order zero-point correction would not sufficiently correct for the discrepancy. On the other hand, synthetic data overlaps substantially with all observed data, ensuring that a single model trained on the synthetic data may satisfactorily estimate photometric redshifts, independently of which survey is used for the inference. 

%%%%%%%%%%%%%%%%%%%%%%%

\subsection{The probabilistic neural network algorithm}
\label{sec:gmm}

The mapping of broad-band filter values to photometric redshift is complicated due to multiple factors. Primarily, degeneracies arise in the redshift prediction due to the presence of multiple clusters in the colour space $\mathbb{C}$. This results in non-Gaussian posteriors for photometric redshifts $p(z_{\rm phot} | c \in \mathbb{C})$, which is typically ignored in $\chi^2$ optimization schemes in template fitting or regression neural networks. In order to capture multi-modal information within the inference pipeline, one may explicitly include a {\it hard clustering} method (like k-means clustering) to separate the whole galaxy sample sub-types and then perform regression analysis within individual subtypes. Alternatively, one can also perform {\it soft clustering} for determining a probability of association of a given datapoint with a specific sub-type, and perform regression without separate training models. The latter approach is employed in \synz: specifically, a Gaussian mixture model coupled with a fully connected neural network for redshift estimation.

A Gaussian mixture function \citep{McLachlan1988} is a linear combination of several Gaussian components, each identified by $i \in {1, 2, ..., m}$, where $m$ is the total number of clusters, as defined in Equation~\eqref{eq:GMM}.

\begin{equation}\label{eq:GMM}
    p(z_k | c_k) = \sum_{i=1}^{m}\pi_i (c_k) \mathcal{N}( z_k| \mu_i (c_k),\sigma_i (c_k))
\end{equation}

For each galaxy with index $k$, the PDF of redshift conditioned on input photometry $p(z_k | c_k)$ is a combination of Gaussian distributions. This Gaussian mixture is characterized by the means $\mu_i(c_k)$ and standard deviations $\sigma_i(c_k)$ of the Gaussian components, and the mixing probability of association of the datapoint with a specific cluster $\pi_i(c_k)$ , satisfying $\sum_{i=1}^{m}\pi_i(c_k) = 1$. While the number of components in the mixture, $m$, is generally pre-specified or independently optimized, the rest of the parameters are learned using an unsupervised clustering algorithm.

%%%%%%%%%%%%%%%%% %%%%%%%%%%%%%%%%% %%%%%%%%%%%%%%%%% 
\subsubsection{Mixture Density Networks}

\label{sec:mdn}
In order to learn the scalars that parameterise the Gaussian mixture model ${\mu(c_k), \sigma(c_k), \pi(c_k)}$, we implement a Mixture Density Network \citep[MDN;][]{bishop1994mixture} by combining a conventional neural network with a mixture density model to predict the conditional probability in Equation~(\ref{eq:GMM}) for the redshift $p(z_k| c_k)$ for given broad band colours $c_k \in \mathbb{C}$. It has to be noted the this training itself is supervised, i.e., the target $z_{\rm spec}$ is provided for every $c_k \in \mathbb{C}$, but the clustering itself is inherently unsupervised.  

A crucial improvement in this prediction model is achieved by replacing the mean-squared error type of penalty (or a loss function), $l(\Theta) \propto |\{z_{\mathrm{phot}}\}(\Theta) - \{z_{\mathrm{spec}}\}|^2$ (equivalent to negative log-likelihood for the case of a single-component Gaussian distribution) as a function of network's model parameters $\Theta$ with a more generic negative log likelihood function given in Equation~\eqref{eq:negloglike}:

\begin{align}\label{eq:negloglike}
    l(\Theta) = -\log \prod_k p(z_k | c_k, \Theta) =  - \sum_k \log p(z_k | c_k, \Theta) \\ = - \sum_k \log \sum_{i=1}^{m}\pi_i (c_k, \Theta) \mathcal{N}( z_k| \mu_i (c_k, \Theta),\sigma_i (c_k, \Theta))
\end{align}

Since the likelihood of the dataset is simply $\prod_k p(z_k | c_k)$, the loss $l(\Theta)$ as a function of network parameters $\Theta$ (weights and biases of the network) reduces to a convenient summation form whose derivative terms are easily obtained during training of the neural network. 
We train the network to maximize the sum of the log-likelihoods of the output PDFs given in Equation~\eqref{eq:negloglike}. 

\begin{figure}
	\includegraphics[width=\columnwidth]{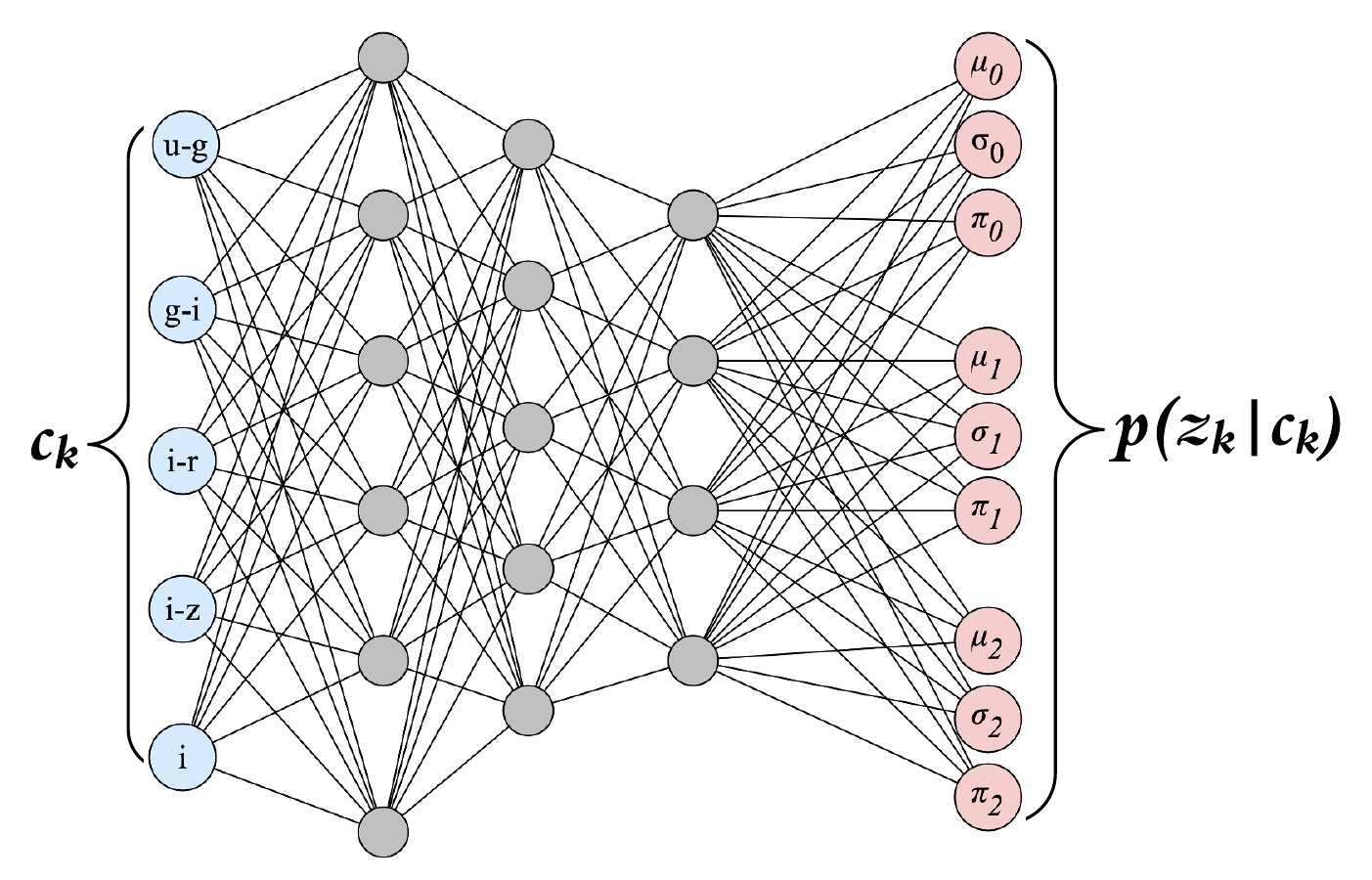}
    \caption{Neural network architecture for the Mixture Density Framework. The colour-magnitude inputs are mapped to the Gaussian mixture model parameters in order to obtain a conditional density distribution $p(z_{\rm phot} | c)$.}
    \label{fig:mdn_arch}
\end{figure}

The prescription of minimizing a mean-squared (or similar point-wise) error function lacks a probabilistic view since one does not explore the likelihood surface around the minimum. Hence error bars on the estimates are typically absent in the outputs of such neural networks. In contrast, the MDN provides an inexpensive approach to quantify the epistemic uncertainties in the predictions, as demonstrated further in Section \ref{sec:uq}. Hence MDN is the primary estimation component of the \synz framework.

%%%%%%%%%%%%%%%%% %%%%%%%%%%%%%%%%% %%%%%%%%%%%%%%%%% 

\subsubsection{Training procedure}
\label{sec:training}

The first step towards training an ML-based prediction model lies in curating the data. We first split the synthetic data into training and `in-set' validation sets. A fraction ($\sim$10\%) of the $\sim$200,000 synthetic datapoints is used for validation and testing purposes. These `hold-out' datapoints are randomly chosen, and do not contribute to the training of the algorithm. With the remaining datapoints, we also perform a re-sampling before training the network. We choose $\sim$200 redshift bins in the range $0.0 < z < 1.2$ and randomly select 50 galaxies in each bin. The resulting re-sampled datapoints are roughly uniform. Since the neural networks are generally sensitive to the underlying distribution of the training data, such bias-mitigation techniques can overcome issues related to imbalanced training schemes. 

Next, we also re-scale both inputs and targets of the neural network model. We standardize the broad band input by removing the mean and scaling to unit variance. The mean and variance are computed across all bands, so that the correlations between the bands are preserved. We then transform the target redshifts by scaling them in the range of 0 to 1. Such pre-processing steps are a common requirement for machine learning tasks. However, the specific choices of the re-scaling transformations can be made heuristically. We have tested between various choices of transformations and settled on the above, based on the rate of convergence and accuracy of the resulting estimations. 

The mixture density network (explained in Section \ref{sec:mdn}) in this implementation is fundamentally a mapping estimator from galaxy colours to redshift. The sample network (Figure \ref{fig:mdn_arch}) is a fully connected dense network with 11 layers with number of neurons per layer as: $[5 (\textrm{input})\rightarrow 512 \rightarrow 1024 \rightarrow 2048 \rightarrow 1024 \rightarrow 512 \rightarrow 256 \rightarrow 128 \rightarrow 64 \rightarrow 32 \rightarrow 1\times 3]$. The last layer [$1\times 3$] corresponds to $(\pi, \mu, \sigma)$ for each grid point that parameterise the predictive conditional distribution $p(z_{\rm phot}|{c})$. Each hidden node in the network has a nonlinear hyperbolic tangent (or a Rectified Linear Unit) activation \citep{activ}. 

The error function in Equation \eqref{eq:negloglike} is optimized during the training, and the weights $\bm w$ are updated using an Adam optimizer \citep{Kingma2014}. The training procedure of the mixture density network is summarized in Algorithm \ref{algo:pnn}. 

\begin{algorithm}
\caption{Training routine for mixture density networks}
\begin{algorithmic}[1]
\label{algo:pnn}
\STATE ${\bm w} \leftarrow $ Initialize network parameters
\WHILE{not converged}
\STATE $x \leftarrow  $ Subset of the complete training data set (a batch)
\FOR{$x_k$ in $x$} 
\STATE Forward propagation i.e., map the input colours to Gaussian mixture model parameters: $(\pi_k, \mu_k, \sigma_k)$ $\leftarrow c_k; {\bm w}$
\STATE Compute predictive distribution: $p(z_{k, {\rm phot}}|c_k) \leftarrow (\pi_k, \mu_k, \sigma_k)$ 
\ENDFOR
\STATE  Compute negative log-likelihood: $\mathcal{E} \leftarrow -\log \mathcal{L}$ 
\STATE Update network parameters using back-propagation: ${\bm w} \leftarrow {\rm Adam}(\nabla_{\bm w} \mathcal{E}, {\bm w})$ 
\ENDWHILE
\end{algorithmic}
\end{algorithm}

The hyper-parameters such as the learning rate, decay rate, and network parameters are optimized heuristically by monitoring the changes in loss (Equation~\eqref{eq:negloglike} with different epochs. However, one may also consider the use of theoretical optimization methods such as hyperopt \citep{Bergstra13} and Bayesian optimization \citep{snoek2012practical} so that the accuracy of the estimation can be improved.

\subsection{Zero-point calibration}

\label{sec:zp}

Adjusting the zero-points of the observational colours to obtain better redshift predictions is a common approach in the template-based photo-z literature~\citep[e.g.][]{Benitez:00,BPZCoe06,cfht_hilde, alhambra, COSMOS2015, PAUSphotoz, PAUSCOSMOSphotoz}. It attempts to correct relative zero-point errors remaining from the zero-point calibration of the data itself, or errors resulting from incorrect point spread function (PSF) modelling ~(\citealt{cfht_hilde}). It will also absorb any leading order systematic offsets between the templates and the data. When the MDN is trained using observations, such zero-point errors are naturally learned by the model connecting redshift with colours. However, since the synthetic data generated to train the network is noiseless, and does not emulate any of the effects imprinted by the observation process, an overall zero-point mismatch is, in general, to be expected.

We parameterise the zero-point correction with four colour offsets for each survey (one for each colour used as input by the MDN). We estimate the zero-point set $\{\lambda_k; k=u-g,g-i,\,i-r,\,r-z\}$ from the measured spectroscopic redshifts $\{z^{j}_{\mathrm{spec}}; j=1\ldots N_{g}\}$ and measured features for a subset of the training sample. The measured features include the measured expected value of the magnitudes and colours, $\{\langle\hat{c}^j_k\rangle\}$ and their measured root-mean-squared values, $\{\sigma(\hat{c}^j_k)\}$. Then, 
\begin{equation} \label{eq:zp_eq1}
\begin{split}
    p(\{\lambda_k\}|\{\langle\hat{c}^j_k\rangle\},\{\sigma(\hat{c}^j_k)\} ,\{z^j_{\mathrm{spec}}\}) =& \prod_{j}^{N_g} p(z^j_{\mathrm{spec}}|\langle\hat{c}^j_k\rangle,\sigma(\hat{c}^j_k),\{\lambda_k\})\, \\ 
    &\times p(\{\lambda_k\}) 
\end{split}
\end{equation}
where we assume all galaxies to be measured independently from each other and note that the zero-points do not explicitly depend on the measured features. The prior $p(\{\lambda_k\})$ is uniform. 
The redshift probability is given by
\begin{equation}\label{eq:zp_eq2}
\begin{split}
      p(z_{\mathrm{spec}}^j|\langle\hat{c}^j_k\rangle,\sigma(\hat{c}^j_k),\{\lambda_k\}) =& \int dc_k \, p(z_{\mathrm{spec}}^j|c_k^j, \langle\hat{c}^j_k\rangle,\sigma(\hat{c}^j_k),\{\lambda_k\}) \\
      &\times p(c_k^j| \langle\hat{c}^j_k\rangle,\sigma(\hat{c}^j_k),\{\lambda_k\}) \\  
      =& \int dc_k \, p(z_{\mathrm{spec}}^j|c_k^j) \\
      &\times p(c_k^j| \langle\hat{c}^j_k\rangle,\sigma(\hat{c}^j_k),\{\lambda_k\})
\end{split}
\end{equation}
where $c_k$ are the noiseless features from the synthetic templates and $ p(z_{\mathrm{spec}}^j|c_k^j)$ is given by the MDN, which only depends explicitly on the synthetic colours (going from the first to the second line in Equation~\eqref{eq:zp_eq2}). The probability of observing the synthetic colours is proportional to the measurement likelihood of each object, which is usually assumed to be a normal distribution given by the measured colour and colour errors. In this case, the normal distribution is centered at the colours plus the zero-point shift,
\begin{equation}\label{eq:zp_eq3}
\begin{split}
      p(c_k^j| \langle\hat{c}^j_k\rangle,\sigma(\hat{c}^j_k),\{\lambda_k\}) \propto  \exp\left(-\frac{\left(c_k^j - (\langle\hat{c}^j_k\rangle+\lambda_k)\right)^2}{2(\sigma(\hat{c}^j_k))^2}\right).
\end{split}
\end{equation}
In Equation~\eqref{eq:zp_eq3} we assume that the magnitude and colours are uncorrelated. Also, note that the zero point for the $i$-band magnitude is fixed at a value of zero (equivalent to assuming a prior distribution of a Dirac delta function centred at zero).

We run an MCMC and keep the zero-point values that maximize the likelihood from Equation~\eqref{eq:zp_eq1}. We calculate Equation~\eqref{eq:zp_eq2} by drawing samples of synthetic colours from $p(c_k^i| \hat{c}_k^i,\{\lambda_k\})$, weighting them by their likelihood at the spectroscopic redshift according to MDN, and then taking the average. Calculating the integral from Equation~\eqref{eq:zp_eq2} exactly can be expensive. An alternative method we follow here is to approximate Equation~\eqref{eq:zp_eq2} by assuming a normal distribution, which we evaluate at the spectroscopic redshift value. We draw samples from $p(c_k^i| \hat{c}_k^i,\{\lambda_k\})$ and calculate the mode of $p(z|c_k^i)$ according to MDN. The mean and variance of the approximate normal distribution are the mean and variance of the set of mode values. We have found the zero-point MCMC to converge much faster with this approximate method than the exact calculation for fainter samples of galaxies, for which a larger integration time is needed to calculate  Equation~\eqref{eq:zp_eq2} exactly. Furthermore, since there can be galaxies with large photometric errors, inaccurate spectroscopic redshift, or some misclassified stars, we clip the minimum value the likelihood of a galaxy can take to prevent these outliers from dominating the overall zero-point likelihood.

\section{Results}
\label{sec:results}

With the development and training of this probabilistic framework, we now present the validation results. Due to the discrepancy between the training data and observational data, we investigate the effect of photometric noise and systematic offsets, and compare the \synz performance against baseline algorithms. The validations in this section are done against SDSS, VIPERS and DEEP2 galaxies, and not with respect to hold-out synthetic data.

\subsection{Effect of calibration}
\label{sec:margin}

\begin{figure}
\includegraphics[width=\columnwidth]{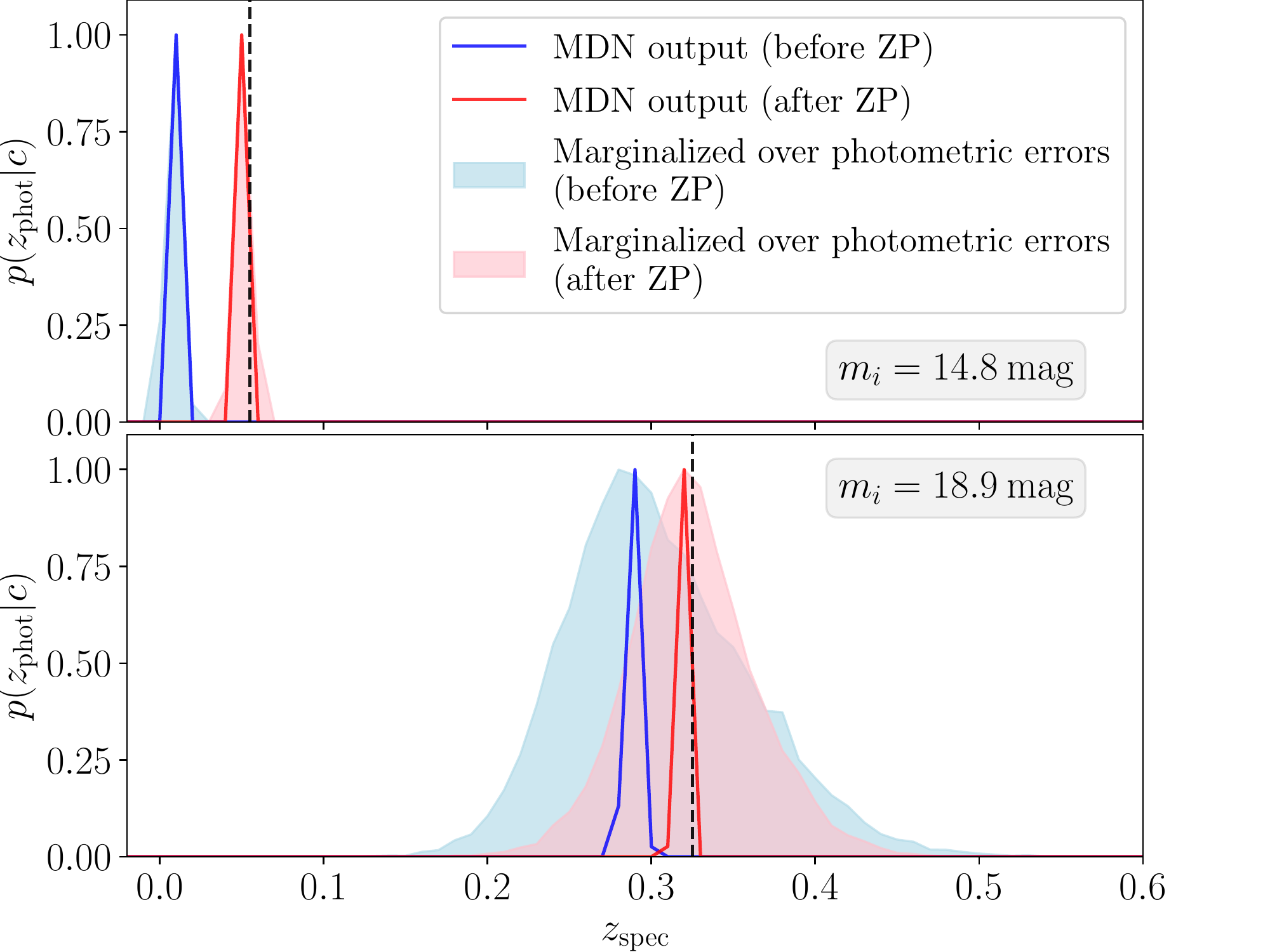}
    \caption{PDFs for a bright, low redshift galaxy (top panel) and a faint galaxy (bottom panel) from SDSS. The $i$-band magnitudes of the galaxies are 14.8 mag and 18.9 mag respectively. The dashed lines show the spectroscopic redshift of the galaxies. The blue solid lines show the MDN output before zero-point (ZP) calibration and the red lines show the post-calibration output. The corresponding shaded regions demonstrate the broadening of the photo-z PDF due to marginalization over observational errors.}
    \label{fig:pdfs}
\end{figure}

The output of the MDN trained from noiseless synthetic data is generally a narrow conditional density estimate $p(z_{\rm{phot}}|c)$. This is expected, since the input colours are point values, while the MDN is trained on tightly sampled noiseless data (as explained in Section \ref{sec:training}). Once the MDN training is complete, we perform a zero-point calibration, as detailed in Section \ref{sec:zp}. The resulting conditional density estimate is also narrow, the motivation for zero-point calibration is to correct for the systematic offsets arising due to noiseless synthetic training data. Both these PDFs are shown for two SDSS galaxies in Figure \ref{fig:pdfs}. The calibration in the colour space results, on average, in the PDFs (the red lines in both panels) agreeing better with the spectroscopic redshift $z_{\rm phot}$ from the SDSS catalogue, than the direct output of MDN before zero-point calibration (the blue lines). 

\begin{figure}
	\includegraphics[width=\columnwidth]{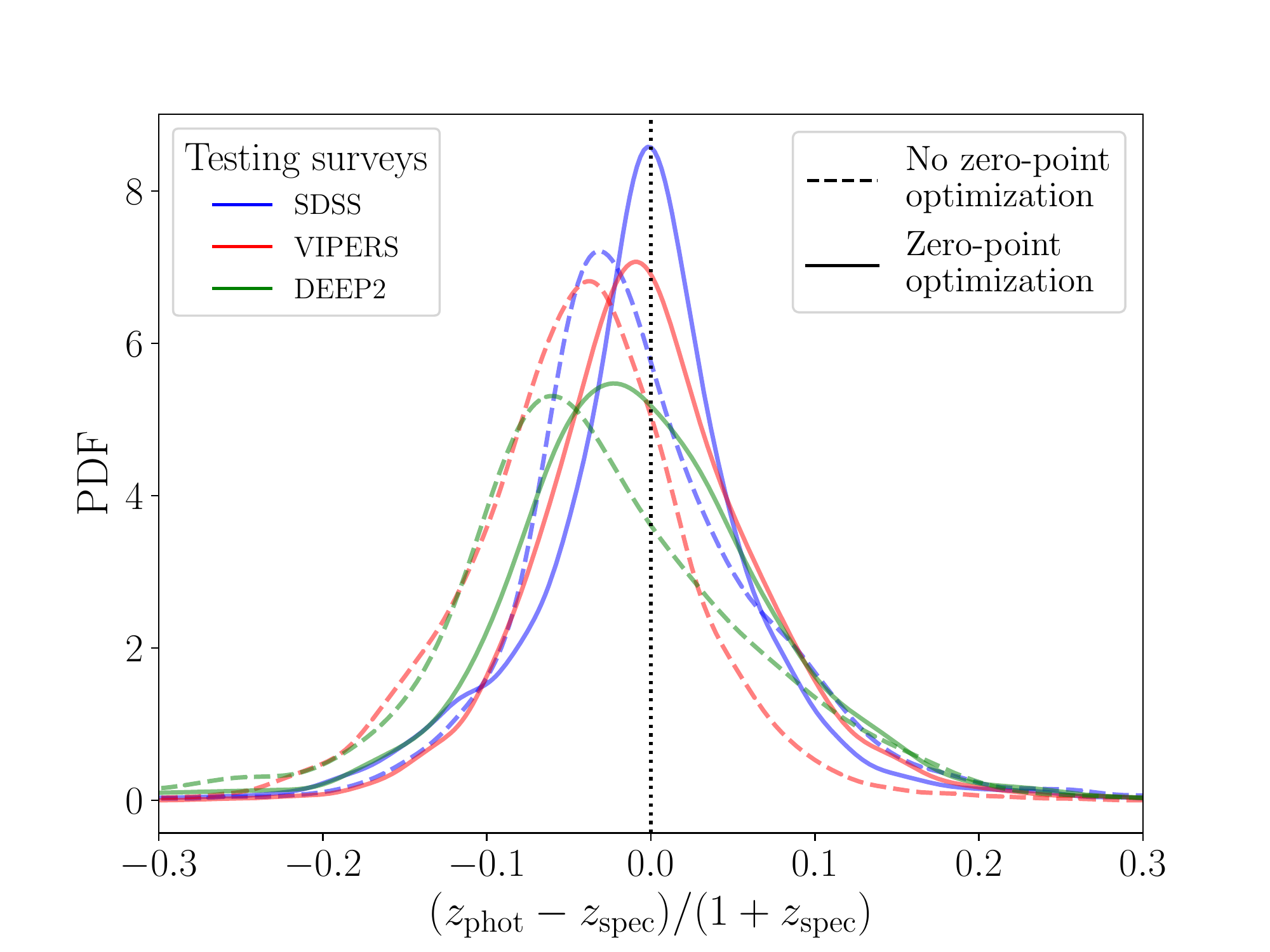}
    \caption{Effect of zero-point correction: The PDF shows the relative difference between the predicted and true redshifts before and after the zero point correction. For each survey, the zero-point optimization (solid lines) improves the accuracy compared to the original MDN results (dashed lines). }
    \label{fig:zp_pdf}
\end{figure}

Next, we demonstrate the effect of the colour calibration process on all three observational surveys of interest. The SDSS, VIPERS, and DEEP2 relative differences $(z_{\rm phot} - z_{\rm spec})/(1 + z_{\rm spec})$ before calibration peak away from zero, as shown in Figure~\ref{fig:zp_pdf}. With the zero-point correction, this distribution peaks closer to zero, signifying improved accuracy in the estimated redshifts. 

\subsection{Uncertainty estimates}
\label{sec:uq}

When the MDN measurements are marginalised over the observational measurement error, the $p(z_{\rm{phot}}|c)$ is broader. We investigate this by running 1000 MDN evaluations per galaxy, where the input colours are drawn from Gaussian distributions corresponding to the SDSS photometric error (\textit{i.e.} we use Equation~\ref{eq:zp_eq2}). Figure~\ref{fig:pdfs} shows the broadening of the photo-z PDF due to averaging over the 1000 conditional density measurements, before (blue shaded region) and after (red shaded lines) the zero-point correction. 

Once the systematic offsets are corrected using the zero-point corrections, the pre-marginalized PDF output of the MDN represents the epistemic uncertainty and is described via the distribution of the weights of the neural network. Epistemic uncertainties are usually  model dependent, i.e., they can be reduced by an optimal training process or a better statistical model. Due to the tight sampling of training data points and near-optimal training of the MDN, the epistemic uncertainty is highly minimized, as demonstrated in Figure~\ref{fig:pdfs}. Next, the PDF obtained by marginalizing over the observational colours (after zero-point correction) represents the aleatoric uncertainty, arising from the inherent colour variability. The stochasticity in the prediction is limited by the quality of the observational photometry, and cannot be improved by a better training scheme or a more flexible neural network. 

\begin{figure}
\includegraphics[width=\columnwidth]{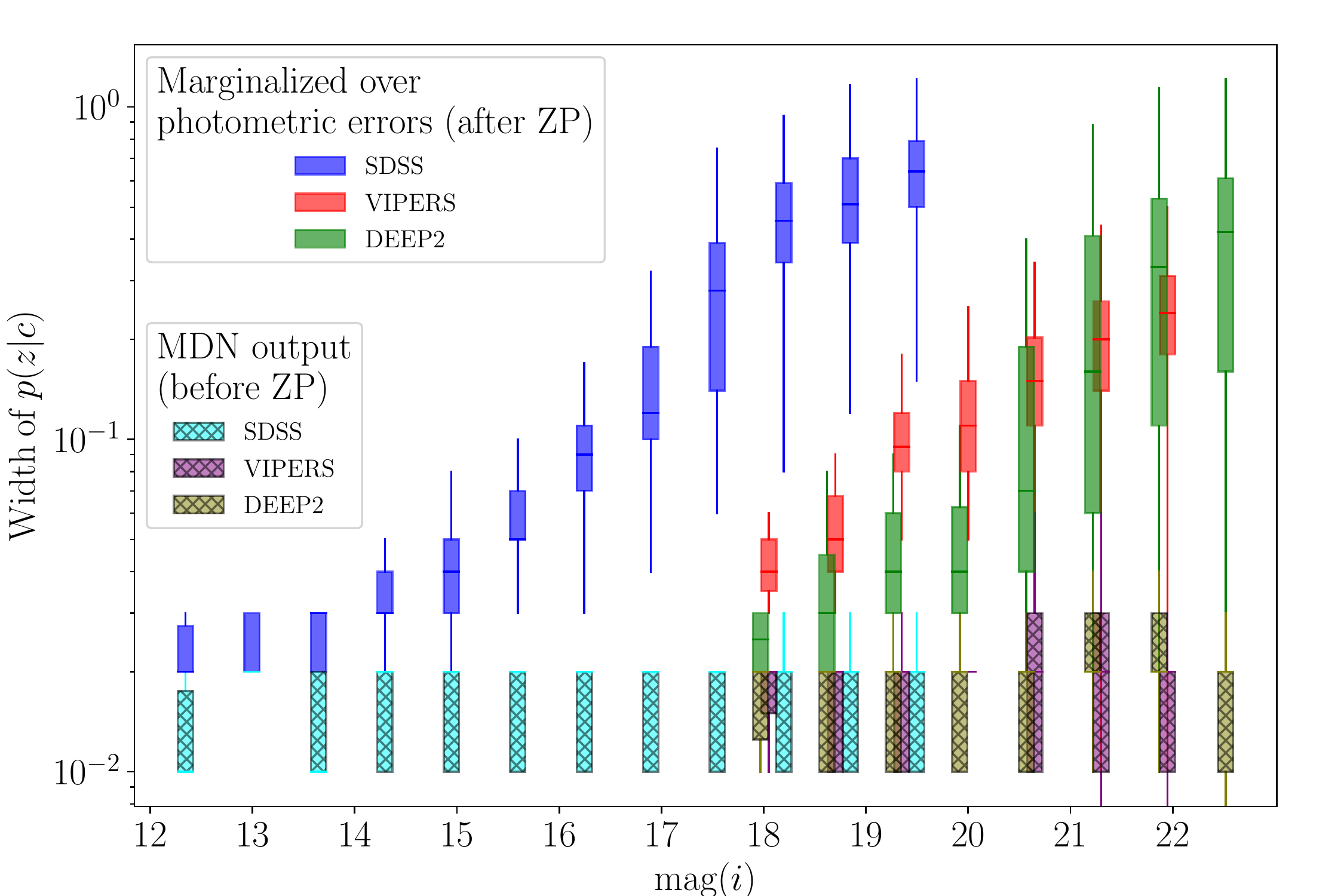}
    \caption{Widths of photo-z PDFs at different $i$ band magnitude bins. The widths are shown for the three observational surveys - SDSS, VIPERS and DEEP2 -- used for testing. The widths are shown for two types of PDFs. The box plots with hatches show the MDN output before zero-point calibration, and the solid box plots are for photo-z PDFs marginalized over observational errors (with zero-point correction).}
    \label{fig:zp_std}
\end{figure}

The two panels in Figure~\ref{fig:pdfs} also show the difference between redshift estimates for a bright, low redshift galaxy and a fainter galaxy at a higher redshift. Marginalization over the observed photometric errors results in a wider broadening for the faint SDSS object in the lower panel. In Figure \ref{fig:zp_std}, the effect of obtaining less confident predictions with fainter redshifts is demonstrated for all the galaxies in our testing sample. For the SDSS, VIPERS and DEEP2 surveys, the width of marginalized $p(z_{\rm phot}|c)$ increases with larger i-band magnitudes. In other words, the aleatoric uncertainty of the redshift prediction increases with increasing magnitude. On the other hand, the corresponding width of the MDN density estimates is both smaller, and roughly consistent across varying brightness of the galaxies. This pre-marginalization $p(z_{\rm phot}|c)$ includes the uncertainty due to the training data prior. That is, the width is primarily due to the limitations in the training process, the information loss due to the small number of broad bands, and their colour-redshift degeneracies in the limited number of bands. However, the MDN output alone does not capture the effect of measurement error in the surveys.   

The process of zero-point calibration implemented within the \synz framework (from Section~\ref{sec:zp}) involves an assessment of model uncertainty using the noisy galaxy samples and a consequent updating of the redshift prediction model. This inverse uncertainty quantification is often neglected in machine learning regression algorithms; most prediction codes instead quantify the forward uncertainty propagation (i.e., the influence on the outputs from the parametric variability in the input space) by assessing the mean, variance, or the distribution of outputs. An alternative to zero point calibration is a bias correction, where a discrepancy function can be determined based on spectroscopic redshift values for a small number of observational data points. Within the current \synz framework, we only perform the parameter calibration. However, the bias correction may be applied independently, or in combination with the calibration.  

Hence, the \synz framework has a two-fold treatment of the uncertainties. First, it provides an inverse assessment of uncertainty and calibration in the input space, to account for the systematic offsets between the trained MDN model and true photometric redshift mapping. Second, it includes a forward uncertainty propagation, identifying the sources of both aleatoric and epistemic uncertainties in the redshift estimates.

\subsection{Comparison with observational training}
\label{sec:compare}

We now compare the novel photometric redshift estimation technique of \synz against baseline schemes where the training is done purely on observational datasets. To ensure uniformity over the comparison models, we use a similar scheme for training while using SDSS and VIPERS data. That is, both SDSS and VIPERS datasets (explained in Section \ref{sec:data_obs}) are divided into training and testing sets. 
 
While our method provides a PDF parameterised by Gaussian mixtures for prediction of redshift, several popular metrics rely on point estimates. In order to get the best redshift point estimate, we choose the mean of the Gaussian component with the highest weight.
%, which is equivalent to the mode of the PDF. 
This procedure is commonly used to reduce photo-z PDFs to a point estimate (see Appendix B of \citealt{Schmidt2020}). The standard deviation corresponding to the highest weighted component is taken as the best Gaussian uncertainty estimate.  

\begin{figure}
    \centering
    \begin{subfigure}[b]{\columnwidth}
        \centering
        \includegraphics[width=\textwidth]{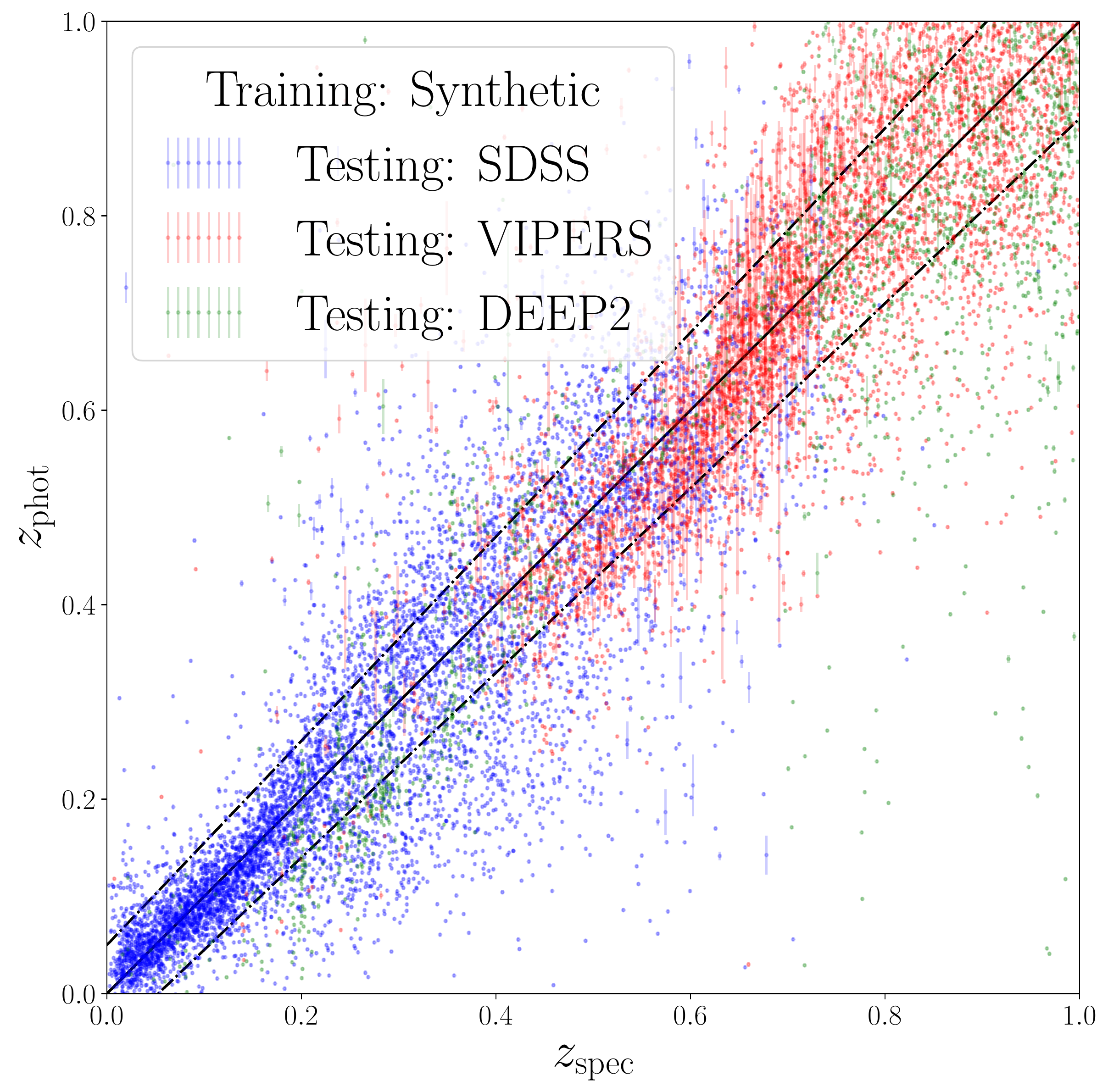}
        \caption[]%
        {{\small Synthetic training (with zero-point correction).}}   
        % \label{fig:mean and std of net14}
    \end{subfigure}
    \vskip\baselineskip
    \begin{subfigure}[b]{0.475\columnwidth}   
        \centering 
        \includegraphics[width=\textwidth]{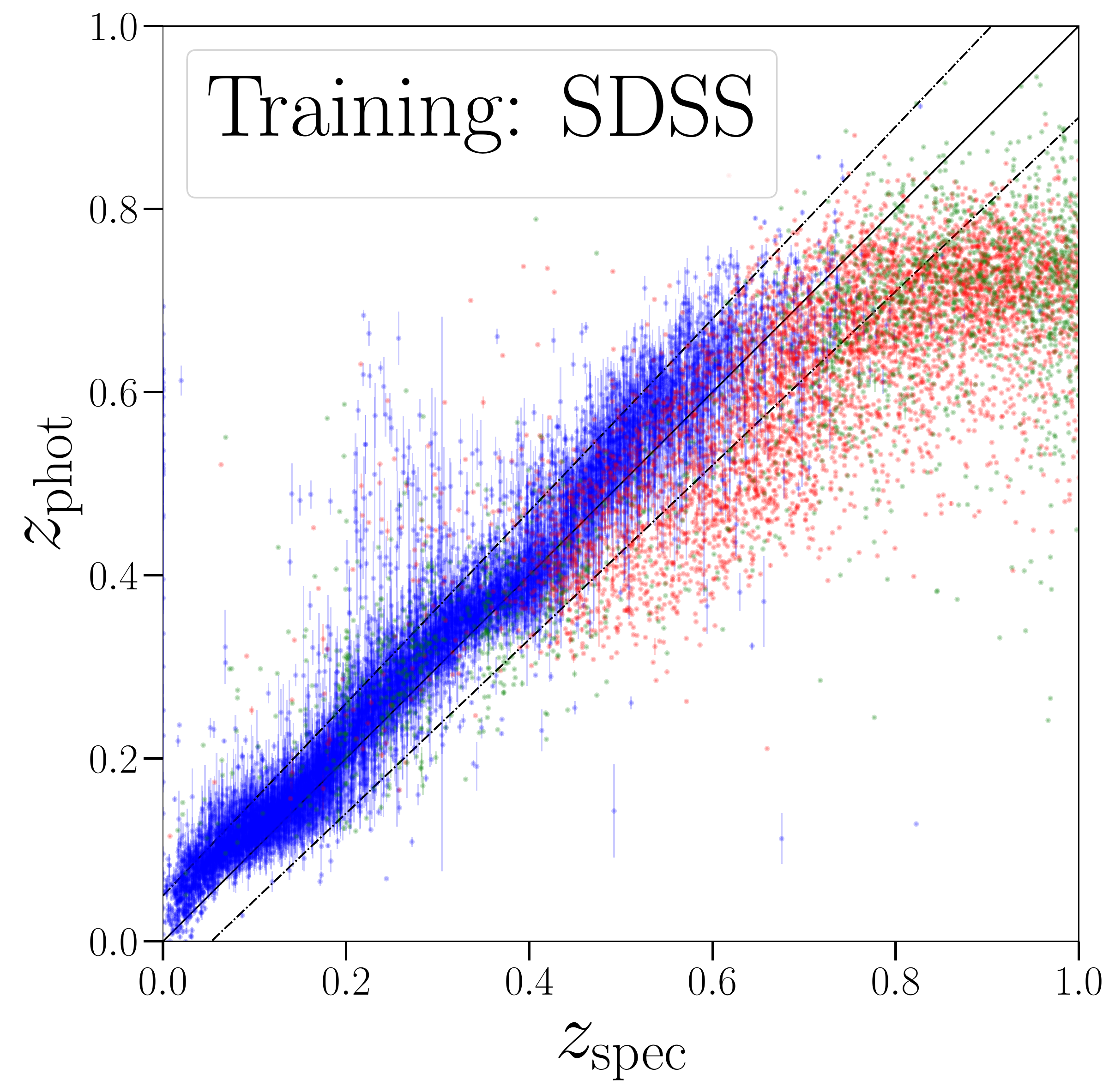}
        \caption[]%
        {{\small SDSS training}}    
        % \label{fig:mean and std of net34}
    \end{subfigure}
    \hfill
    \begin{subfigure}[b]{0.475\columnwidth}  
        \centering 
        \includegraphics[width=\textwidth]{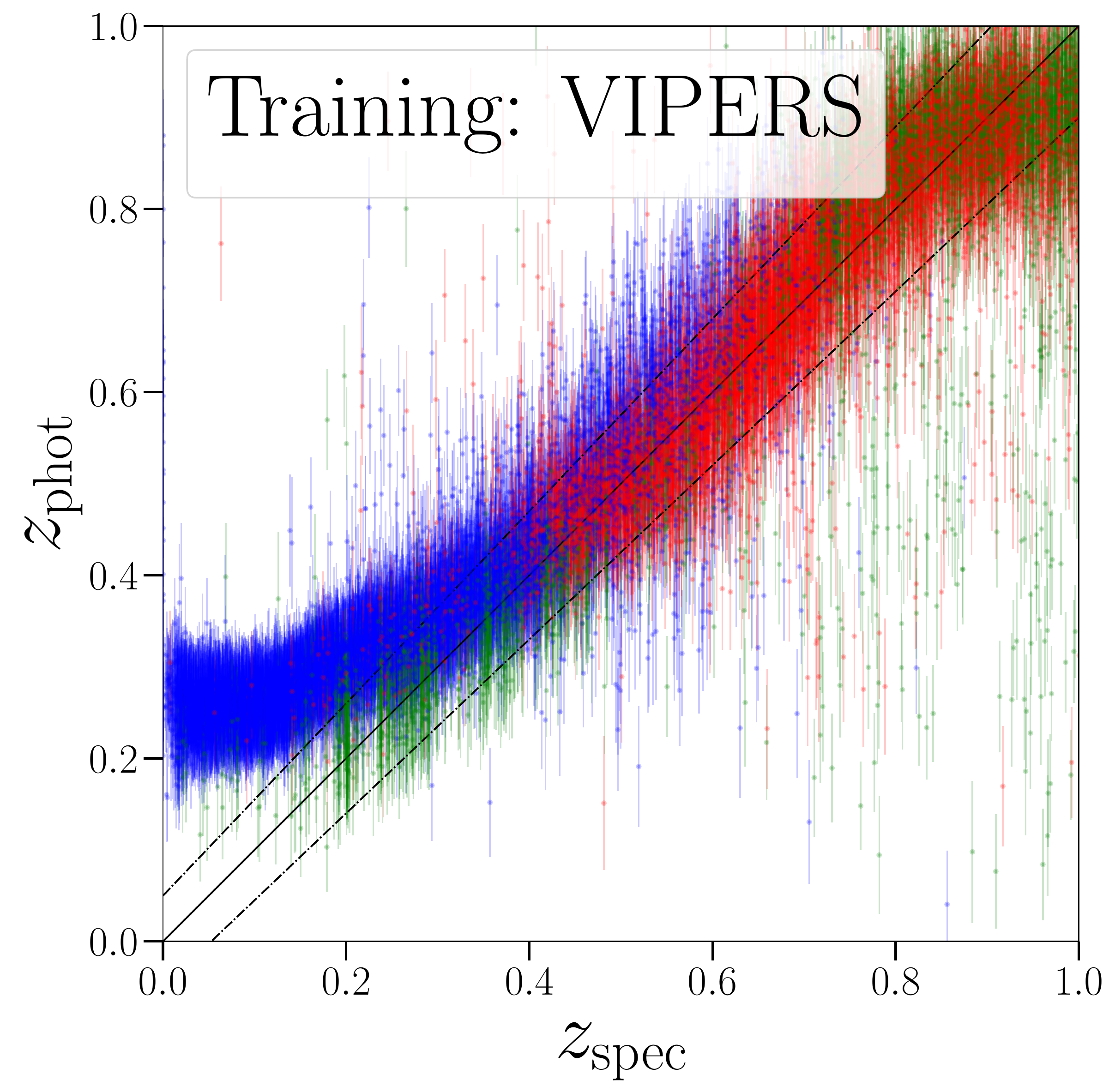}
        \caption[]%
        {{\small VIPERS training}}    
        % \label{fig:mean and std of net24}
    \end{subfigure}

    \caption[]
        {\small Comparison of predicted photometric redshifts and spectroscopic redshifts for observed galaxies. The training is performed using synthetic data only (top panel), VIPERS (bottom left) and SDSS (bottom right), respectively. All the error bars shown in the plots correspond to MDN outputs without marginalizing over the photometric errors.} 
    \label{fig:all_compare}
\end{figure}

For comparisons, we use the same observational test set for all three training schemes. Photometric redshift estimates from the three training schemes are shown in Figure~\ref{fig:all_compare}. The top panel shows the results from the synthetic colour-trained \synz framework, where the estimates for SDSS, DEEP2 and VIPERS show reasonable agreement with the spectroscopic redshifts. In comparison, an SDSS-trained model in the lower left panel performs well with the test SDSS dataset, but the estimations for VIPERS and DEEP2 are significantly biased. This bias arises because SDSS galaxies are selected to be preferentially red at intermediate and high redshifts, while DEEP2 and VIPERS contain mostly blue galaxies (Figure~\ref{fig:data_colour_comparison}). A similar trend is seen when an ML model is trained with VIPERS (lower right panel of Figure~\ref{fig:all_compare}): the estimation for hold-out VIPERS and DEEP2 data sets are both unbiased, whereas redshifts are under-estimated for SDSS colours.

Notable differences across different training schemes can be seen in the standard deviations of the redshift predictions $p(z_{k, {\rm pred}}|c_k)$, depicted by the error bars in Figure~\ref{fig:all_compare}. The conditional density estimate of the MDN is conditioned on the training prior. In the case of synthetic training, the training data is noiseless and densely samples the colour-redshift space. Hence the upper panel of Figure~\ref{fig:all_compare} shows small error bars for predictions for all three noisy observational datasets. This was obtained by using the pre-marginalization PDF (after zero-point calibration), hence only the epistemic uncertainties are captured. By sampling the PDFs based on the photometric errors in the observational data, the conditional density estimation is much broader, as shown in Figure~\ref{fig:zp_std}. On the other hand, the lower panels show broader error bars, since the training data sparsely sample the colour-redshift space. They also capture the photometric error in the observations within the training scheme. This is a crucial difference in the uncertainty quantification of synthetic- and observational- trained models. While with \synz, we are able to isolate epistemic and aleatoric uncertainties simply by pre- and post-marginalization stages, it is simply not possible when the training data is noisy. The error bars of SDSS- and VIPERS- trained models in Figure~\ref{fig:zp_std} include epistemic uncertainty (like the pre-marginalized \synz PDF) and a part of aleatoric uncertainty (unlike the pre-marginalized \synz PDF), without a clear interpretation of their contribution to the overall uncertainty of the MDN. An additional marginalization over photometric errors during the testing phase would provide another contribution to aleatoric uncertainty of the SDSS- and VIPERS- trained model. 

The aleatoric uncertainty in the redshift predictions depends on the internal randomness in the inputs, i.e., the photometric errors. This can not be improved for a given survey. However, the epistemic uncertainty in the redshift predictions can be improved easily in a synthetic-training scheme, plainly by increasing the size of simulated SED templates. This is not a viable option in models trained on observational data, where the available galactic spectra are limited. This limitation is more pronounced for fainter galaxies at high redshift, where the spectroscopic follow-up is prohibitively expensive. 

In addition, a neural network trained on SDSS may yield less-explainable error bars when tested on a survey with a different error model, such as VIPERS. Hence, understanding the uncertainty propagation and interpreting the error bars with estimation techniques like neural networks can be relatively more difficult in noisy training schemes.

\subsection{Aggregate metrics for performance evaluation}
\label{sec:metrics}

\begin{figure*}
	\includegraphics[width=\textwidth]{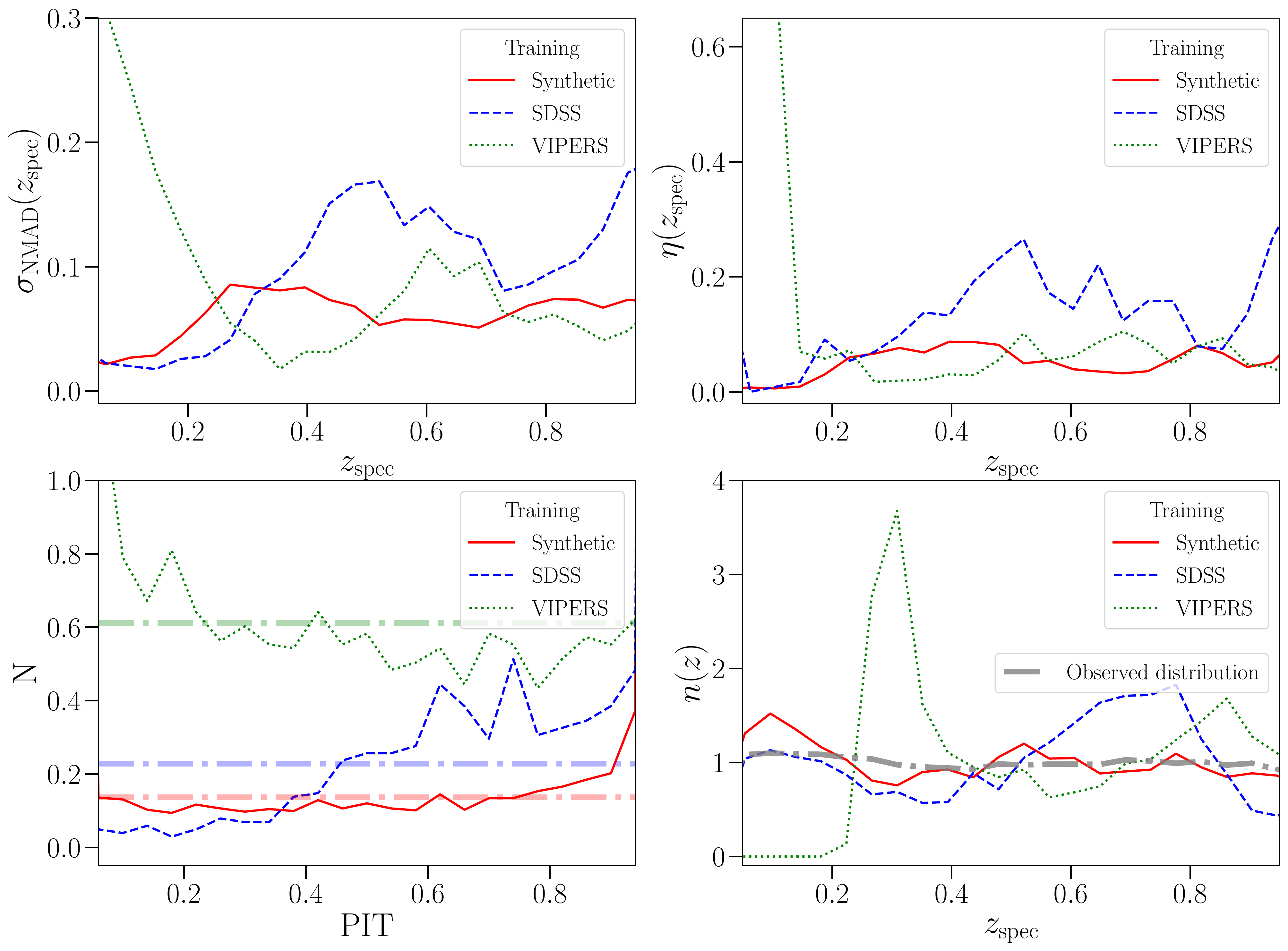}
    \caption{Comparisons between 3 different training schemes, i.e., VIPERS-, SDSS- and synthetically-based training. Testing is done on a combined observational dataset that is held-off from all 3 training schemes. Metrics used are the normalized median absolute deviation $\sigma_{\rm NMAD}$ (top left panel), The catastrophic outlier fraction $\eta$ (top right panel), Probability integral transform (bottom left panel) and Galaxy redshift distribution $n(z)$ (bottom right panel). The shaded dash-dotted lines in the bottom left panel show the PIT curves for perfect estimations. The shaded region of the bottom right panel shows the Poisson errors in the bin-counts of the $n(z)$ distribution. }
    \label{fig:metrics}
\end{figure*}

In order to quantitatively compare the performance of our synthetically-trained photometric redshift estimation model \synz with baseline comparison methods using observational training, we consider a few commonly-employed metrics. These benchmarks utilize redshift estimates without a marginalization over photometric errors. 

\begin{enumerate}

  \item \textbf{Normalized median absolute deviation:} As a measure of the accuracy of the model, we use the Normalized median absolute deviation $\sigma_{\rm NMAD}$ (\citealt{Ilbert2009}) to measure photo-z precision. 
  
  \begin{equation}
  \centering
    \sigma_{\rm NMAD} = 1.48 \times \mathrm{Median}\left( \frac{|z_{\rm spec} - z_{\rm phot}|}{1 + z_{\rm spec}} \right)
\end{equation}

\noindent The scale factor of $1.48$ is used for re-interpreting $\sigma_{\rm NMAD}$ as the standard deviation of normally distributed data. The median of the relative difference is taken because it is less sensitive to extreme values. Here, $z_{\rm phot}$ is taken as the mean of the highest weighted Gaussian mixture component.  

\item \textbf{Catastrophic Outlier fraction:} The fraction of galaxies that satisfy

 \begin{equation}
 \frac{|z_{\mathrm{phot}} - z_{\mathrm{spec}}|}{(1 + z_{\mathrm{spec}})} > f_{\mathrm{cut}}
\label{eq:outlier}
\end{equation}

\noindent are regarded as catastrophic failures. The Catastrophic Outlier fraction $\eta$ is used to quantify the number of failed photometric redshift estimations (defined by $f_{\mathrm{cut}}$).  

\item \textbf{Probability integral transform}: The cumulative distribution function (CDF) of each source evaluated at $z_{\mathrm{spec}}$ is defined as the Probability integral transform (PIT). 

\begin{equation}
    \mathrm{PIT} = \mathrm{CDF}[p, z_{\mathrm{spec}}] = \int_{-\infty}^{z_{\mathrm{spec}}} \mathrm{p}(z){\rm d}z,
\end{equation}
    
\noindent where $z_{\mathrm{spec}}$ is the true redshift of the source $i$, and $\mathrm{p}(z)$ is the probability distribution function, i.e., the 3-component Gaussian mixture model. The integral is computed from $-\infty$ because the $\mathrm{p}(z)$ can have non-zero values for negative $z$ values. The CDF values at the true redshift over the redshift predictions indicate the over-dispersion, under-dispersion and the biases. The PIT histogram for accurate predictions is uniform between 0 and 1, and the overestimated and underestimated uncertainties are concave and convex respectively.  

\item \textbf{Galaxy redshift distribution}. The overall redshift distribution of the sample, $n(z)$ provides a cumulative metric for accuracy of the photometric redshift estimation. The point redshift estimate $z_{\mathrm{phot}}$ of the MDN is chosen based on the weights of the individual Gaussian components, similar to the redshift estimations used in $\sigma_{\rm NMAD}(z_{\mathrm{spec}})$ and $\eta(z_{\mathrm{spec}})$. However, we note one could use a more appropriate $n(z)$ estimation (using approaches similar to \citealt{Malz2020}) using the probabilistic estimate $p(z_{\rm pred}|c)$ for individual galaxies.
  
\end{enumerate}

We expect the performance of the photometric redshift estimation scheme to change as a function of redshift. Therefore, we explore the above metrics in redshift bins, as shown in Figure~\ref{fig:metrics}. The redshift dependence of metrics such as $\sigma_{\rm NMAD}$ or $\eta(z_{\mathrm{spec}})$ are more indicative of the performance of neural networks, rather than an average over all the testing samples.  

% sigma_nmad

The top left panel of Figure~\ref{fig:metrics} shows $\sigma_{\rm NMAD}$ variation with spectroscopic redshift. The synthetically-trained model value lies consistently under  $\sigma_{\rm NMAD}(z_{\mathrm{spec}}) < 0.1$. On the other hand, the SDSS-trained model has lower $\sigma_{\rm NMAD}(z_{\mathrm{phot}})$ for $z_{\mathrm{spec}} < 0.3$, but rises over $\sigma_{\rm NMAD}(z_{\mathrm{spec}}) > 0.1$ at higher redshifts. Similarly, at low redshifts ($z_{\mathrm{spec}} < 0.25$) the VIPERS-trained model has higher $\sigma_{\rm NMAD}(z_{\mathrm{phot}})$ than that of the synthetic model.

% outlier

As in the case with $\sigma_{\rm NMAD}(z_{\mathrm{spec}})$, a point redshift estimate $z_{\mathrm{phot}}$ of the MDN is chosen based on the weights of the individual Gaussian components. In our comparisons, the cutoff for the failures is chosen to be $f_{cut} = 0.15$. While the VIPERS-trained model has more catastrophic outliers (up to 60\%) at $z_{\mathrm{spec}} < 0.1$, the SDSS-trained model has a large number of outliers in the range $0.5 < z_{\mathrm{spec}} < 0.8$. The synthetically-trained model performs well across the entire range of $z_{\mathrm{spec}}$, with the outlier fraction $\eta(z_{\mathrm{spec}})$ being consistently under 10\%. 

% PIT
In the bottom left panel of Figure~\ref{fig:metrics}, we show the PIT curve in the absence of the estimation outliers by discarding the ${\rm PIT} < 0.05$ and ${\rm PIT} >  0.95$. Both SDSS- and VIPERS-training show significantly more biased, and under-dispersed redshift-estimation PDFs. This is seen in the deviation from the PIT curves corresponding to perfect estimations (shown in dash-dotted lines). The synthetic-training yields in a PIT curve that is closest to a uniform distribution, showing a better-calibrated redshift estimation. 

%  n(z) 
As a final comparison, the bottom right panel of Figure~\ref{fig:metrics} shows a comparison of $n(z)$ for the three training schemes, along with the standard deviation of the galaxy counts per bin. The VIPERS-trained photometric redshift distribution deviated from that of the observed sample (which is the `hold-out' sample of SDSS and VIPERS galaxies) at low redshifts. Similarly the SDSS-trained sample deviates significantly at $z_{\mathrm{spec}} > 0.5$. 

With these aggregate comparisons, the performance of \synz is shown to be superior to that of the baseline models trained on observational data points. While the metrics are by no means exhaustive, they are sufficient to demonstrate the advantages of using realistic synthetic data in photo-z modelling. A more complete analysis with performance metrics, in the context of future surveys like LSST or Euclid, is reserved for later investigations. 

%%%%%%%%%%%%%%%%% %%%%%%%%%%%%%%%%% %%%%%%%%%%%%%%%%% 

\section{Discussions and Conclusions} \label{sec:discussion}

We have presented a hybrid framework for probabilistic photometric redshift estimation, \synz, with 3 notable differences from existing photo-z codes:

\begin{itemize}
    \item The training set comprises entirely of synthetic photometry generated from \fsps and star formation histories from realistic galaxy formation models (as shown in Section \ref{sec:data_syn}). 
    \item A probabilistic network using Gaussian mixture modelling (described in Section \ref{sec:gmm}), and marginalization over noisy observational photometry isolates the aleatoric and epistemic uncertainties in the redshift prediction (explained in Section \ref{sec:uq}). 
    \item Survey-specific calibration is provided to mitigate the leading order offsets between noiseless synthetic photometry and noisy observations (formulated in Section \ref{sec:zp} and the results shown in Section \ref{sec:margin}.)  
    
\end{itemize}

Development of such a synthetic-data based photometric redshift estimation framework requires meticulous experimental design, where both modelling of synthetic SEDs (shown in Figure \ref{fig:templates}) and their colour-redshift distribution (in Figure \ref{fig:data_colour_comparison}) has to match with the observational surveys. Once a surrogate model for mapping photometry to redshifts is trained, systematic offsets in the prediction can be corrected (Figure \ref{fig:zp_pdf}) via a calibration technique inspired by template-fitting algorithms. 

Utilizing this framework, we have demonstrated, using Figure \ref{fig:metrics}, that for a choice of inference method with uncertainty quantification, physical modelling of synthetic data can outperform purely observation-based training. This work motivates the development of robustness requirements for synthetic forward modelling efforts and error propagation modelling in the context of upcoming astronomical surveys.

The spectroscopic observational campaigns are expensive, which results in restricted coverage over sky area or over faint objects. As a result, such samples are incomplete and prone to selection effects, which lead to biased training for ML techniques. On the other hand, template-based methods usually rely on simple libraries, as physically realistic galaxy SED models quickly become too computationally costly. In this paper, we provide a bridge between these worlds by using a set of realistic synthetic templates based on SFHs from state-of-art hydrodynamical simulations and empirical galaxy formation models, together with a probabilistic neural network framework to estimate fast galaxy redshifts conditioned on LSST-like photometry.

Each individual component of \synz (in Section \ref{sec:modeling} can be replaced, say, by a more realistic template library, with a better ML algorithm, or a higher-order offset mitigation technique. However, the principles of designing such a framework -- the meticulous experimental design, a physical forward model for synthesizing galaxy colours, usage of uncertainty quantifiable ML models, and marginalization of estimates over stochastic measurements of observational colours -- remain the same. 

Investigation into synthetic galaxy models, simulations and realistic galaxy mock catalogues are crucial for astronomical studies in the near future. Stage IV dark energy surveys like Euclid and Roman Space Telescope would require about 5000 spectroscopic observations to enable photometric calibrations \citep{Stanford2021}. Training and calibration for surveys like the Rubin Observatory's LSST may require an even higher number (about 30,000) of spectroscopic objects over roughly 15 widely separated regions~\citep{Newman2015}. Larger, more complete samples reaching the faintest objects would be required to reduce the scatter in photometric redshift estimates. Our approach of synthetic data creation circumvents the need for the highly expensive spectroscopic follow-up surveys needed for photo-z calibration. In addition, it also avoids the issue of training with existing observational data that are neither representative nor complete.

In addition, quantification of error propagation is increasingly important for future sky surveys that probe faint galaxies with broad photometric bands. Degeneracies resulting from the limited number of filters results in both biased estimations of redshift, and catastrophic outliers in the predictions. With the Gaussian mixture model we have employed in \synz, followed by the marginalization over observational errors, we are able to model complex photo-z PDFs, two of which are shown in Figure \ref{fig:pdfs}. Incorporating the mixture models not only allows for modelling arbitrary prediction posteriors that are not simple Gaussian distributions but also accounts for degeneracy in the colour-redshift space. 

The possible offsets in the predictions of the MDN can be traced to the lack of modelling of the observational processes, such as the PSF response. This offset is specific to synthetic training models, and calibration in the input space is necessary to ensure an unbiased redshift estimation via the \synz framework. Stringent requirement of unbiased redshift estimation in weak lensing studies \citep{Ma2006} necessitates such an offset-mitigation treatment. 

Finally, the computational expense of photometric redshift estimation is also worth noting. Classical template fitting codes can take over 1 second per object. In comparison, MDN prediction takes less than 0.02 milliseconds per galaxy with a single NVIDIA V100 Tensor Core GPU. This speed-up is reached because the input colours can be passed to the trained network in arbitrarily large batches (only limited by the device memory). In the context of LSST and Euclid, where photometric data of billions of galaxies are expected, this level of robustness is required for most photometric analyses.

Next, we discuss the limitations of the \synz technique, which are applicable to a broader family of photometric redshift estimation codes that rely on synthetic SED templates: 

\begin{itemize}

\item One possible shortcoming of our synthetic data creation is that the \umachine or \illustris models may not generate SFHs representative enough for all galaxies from the observations. To solve this problem, new approaches to model the galaxy-halo connection that brings together the advantages of SFH parametric and empirical models (Alarcon et al. 2021, in prep.) are currently being investigated. Furthermore, the development of surrogate SPS models based on such empirical models to speed up the generation of galaxy photometry (Hearin et al. 2021, in prep.) is also underway. 

\item Another notable issue with approaches that may use synthetic data is that the template library may not be representative of the galaxies from the observations. That is, the templates may not be compatible with any survey galaxy SED. We believe our approach of using ML circumvents this issue by learning the overall mapping, rather than the individual one-on-one matching of galaxies. Regardless, the validity of the synthetic templates used in \synz is ensured only up to redshift of $z_{\rm spec} \approx 1$ (see Appendix \ref{app:tempaltes}). Extensive benchmarking of synthetic templates will be necessary to broaden the scope of \synz, where proper scoring rules or other evaluation metrics would be required.

\item In addition, one should also ensure that the synthetic-trained ML model does not extrapolate in the colour-redshift space of the observational surveys. The black-box nature of over-parameterised statistical models (e.g., deep neural networks) renders the interpretability of the mapping exceedingly difficult whilst extrapolating. Within this study, we have only applied limiting constraints on our data to ensure overlap between synthetic and observed colours. However, agreement of training and testing data distributions in a higher-dimensional space needs to be investigated.  

\item Finally, our treatment for dealing with photometric errors involves large numbers of draws in the colour space, followed by averaging over PDFs from the MDN. The plain estimations of MDN (with zero-point calibration) are much faster, and provide mean redshift predictions with associated epistemic uncertainties, but not the effect of observational errors. The effect of aleatoric uncertainty becomes increasingly important in fainter galaxies. This trade-off between inference speed and accounting for all the prediction errors can be challenging while scaling the model to billions of galaxies. We are currently exploring models based on these requirements, including adaptability to high-performance computing resources, and ML techniques of transfer learning and Bayesian neural networks. 

\end{itemize}

This work facilitates exploration into multiple research avenues. In this work, we have only discussed the synthetic data corresponding to 5 (broad) bands. Detailed analyses with various numbers of bands (for example, 5 LSST colours, 40 narrow bands of PAUS, 56 narrow bands from J-PAS, or 101 SPHEREx channels) are easily possible. Investigations of correlations between individual bands and their effects on photometric redshift estimation can help in designing future surveys. Second, ML inference algorithms can be designed specifically for cross-matched catalogues using synthetic data. In real observational surveys, obtaining a pristine cross-matched sample depends on sky coverage, resolution, and sensitivity of individual surveys. This is prone to multiple cross-matches for single sources and mismatches. However, the synthetic SEDs will only have to be convolved with filter transmission curves of different telescopes in order to get perfectly matched galaxies.  Third, synthetic colours also enable the possibility of inferring star formation rates, metallicities and other galactic parameters. A nonlinear mapping between these galactic properties and photometric colours can be machine-learned since the synthetic data involves these parameters as inputs. Such synthetic forward models are highly valuable if the computational expense is reduced considerably. The above analyses are under active development and will be presented in the near future.

%%%%%%%%%%%%%%%%% %%%%%%%%%%%%%%%%% %%%%%%%%%%%%%%%%% 

\section*{Acknowledgements}

We thank Andrew Hearin for his useful suggestions and for generating \umachine data. We thank Alba Vidal-Garc\'ia, Dave Higdon, Earl Lawrence and Markus Michael Rau for their useful discussions. SH acknowledges an early motivating discussion with Alex Malz.

Work at Argonne National Laboratory was supported by the U.S. Department of Energy, Office of High Energy Physics. Argonne, a U.S. Department of Energy Office of Science Laboratory, is operated by UChicago Argonne LLC under contract no. DE-AC02-06CH11357. This material is also based upon work supported by the U.S. Department of Energy, Office of Science, Office of Advanced Scientific Computing Research and Office of High Energy Physics, Scientific Discovery through Advanced Computing (SciDAC) program.

Funding for the Sloan Digital Sky Survey (SDSS) has been provided by the Alfred P. Sloan Foundation, the Participating Institutions, the National Aeronautics and Space Administration, the National Science Foundation, the U.S. Department of Energy, the Japanese Monbukagakusho, and the Max Planck Society. The SDSS Web site is \url{http://www.sdss.org/}. The SDSS is managed by the Astrophysical Research Consortium (ARC) for the Participating Institutions. The Participating Institutions are The University of Chicago, Fermilab, the Institute for Advanced Study, the Japan Participation Group, The Johns Hopkins University, Los Alamos National Laboratory, the Max-Planck-Institute for Astronomy (MPIA), the Max-Planck-Institute for Astrophysics (MPA), New Mexico State University, University of Pittsburgh, Princeton University, the United States Naval Observatory, and the University of Washington.

The \synz is built using the following Python packages:  \verb|TensorFlow| \citep{tensorflow}, \verb|TensorFlow-Probability| \citep{tfp} and \verb|Scikit-Learn| \citep{scikit}. The analyses performed in this paper utilize the following:  \verb|Numpy| \citep{numpyscipy}, \verb|Matplotlib| \citep{matplotlib} and \verb|GetDist| \citep{getdist}.

\section*{Data Availability}

The photometric redshift estimation codes and analyses in this paper are publicly available at \url{https://github.com/nesar/MDN_phoZ/}. Training and test data will be made available upon request. 

%%%%%%%%%%%%%%%%%%%%%%%%%%%%%%%%%%%%%%%%%%%%%%%%%%

%%%%%%%%%%%%%%%%%%%% REFERENCES %%%%%%%%%%%%%%%%%%

\bibliographystyle{mnras}
\bibliography{mybib}
%%%%%%%%%%%%%%%%% APPENDICES %%%%%%%%%%%%%%%%%%%%%
\appendix

\section{The precision of FSPS spectra}
\label{app:tempaltes}

\begin{figure*}
	\includegraphics[width=\columnwidth]{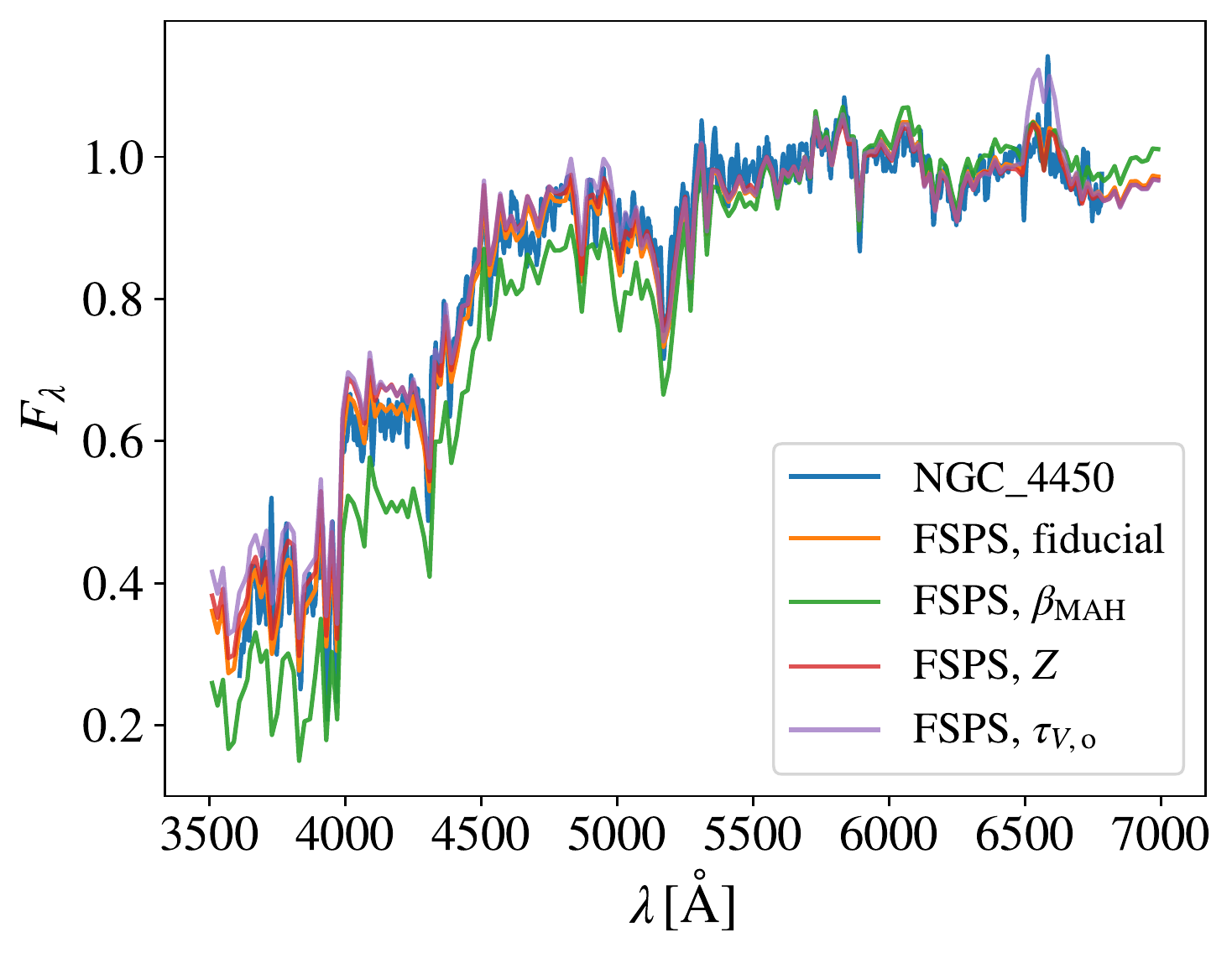}
	\includegraphics[width=\columnwidth]{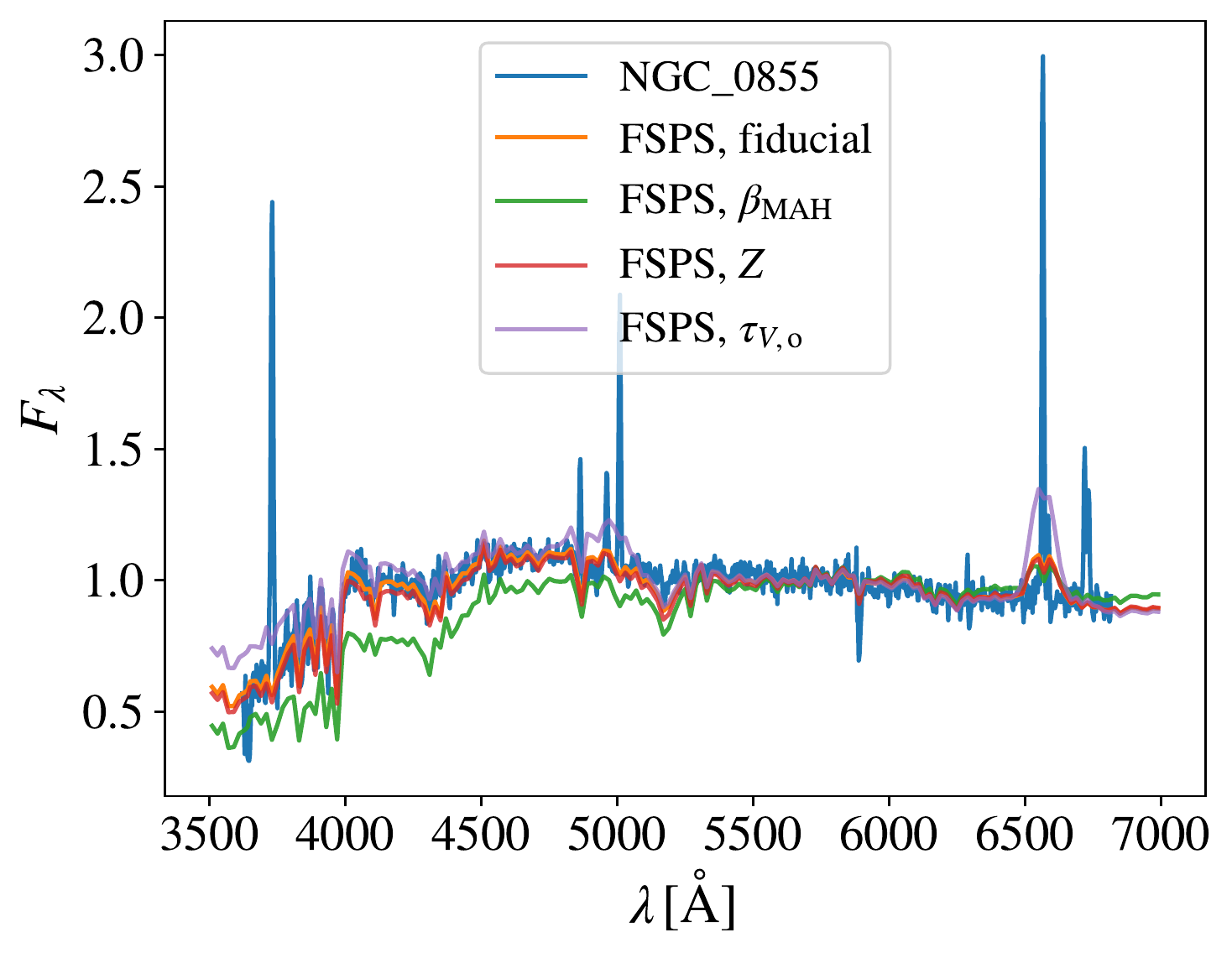}
    \caption{Comparison of observed (blue) and synthetic spectra (orange, green, red, and purple). Left panel: Observed spectra from a quenched galaxy in \citet{Brown2014} compared with the closest fitting synthetic spectra (orange) generated from FSPS. The spectra obtained by perturbing the slope of the mass accretion history $\mahslope$, stellar and gas metallicity $Z$, and the dust attenuation of old stars $\tau_{V}^{\rm ISM}$ from those of the best fitting-template are shown in green, red and purple respectively. Right panel: Similar spectra as the left panel, but for a star forming galaxy.} 
    \label{fig:templates}
\end{figure*}

In this appendix, we show that the synthetic SEDs produced following the methodology described in Section \ref{sec:data_syn} are indeed representative of SEDs from observed galaxies. Figure~\ref{fig:templates} shows the comparison between synthetic and observed spectra. In the left and right panels, blue lines depict the spectra of a randomly selected quenched and star-forming nearby galaxy~\cite{Brown2014}, respectively. Orange lines show synthetic spectra that are representative of the spectra of the previous galaxies. Even though we selected these spectra based on visual inspection, we can readily see that there is a reasonable match between observed and synthetic spectra. 

It is important to note that the resolution of synthetic spectra is worse than that of observed spectra, which explains the smoothing of spectral features like emission and absorption lines. We do not produce spectra with a higher resolution because we are only interested in broad-band colours throughout this work, and the result of convolving high- and low-resolution spectra with broad-band transmission curves is essentially the same.

In Figure~\ref{fig:templates}, we also explore the sensitivity of synthetic spectra to the parameters of our model. To do so, we perturb one of the parameters corresponding to the best-fitting model while holding the others fixed to the best-fitting value. We can see that the largest deviation between the unperturbed and perturbed spectra arises when varying the slope of the host halo mass accretion history, \mahslope, which is indicated by a green line. In comparison, the stellar and gas metallicity $Z$, and the attenuation of old stars (parameterised as given in \citealt{Charlot2000}) present smaller impacts on the spectra.

%%%%%%%%%%%%%%%%%%%%%%%%%%%%%%%%%%%%%%%%%%%%%%%%%%
\bsp	% typesetting comment
\label{lastpage}
\end{document}